\documentclass[preprint,12pt]{elsarticle}

\usepackage{hyperref}
\usepackage{amsmath}
\usepackage{feynmp}
\usepackage{slashed}
\usepackage{enumitem}
\usepackage{placeins}
\usepackage{fancyvrb}
\renewcommand{\i}{\mathrm{i}}
\newcommand{\e}{\mathrm{e}}
\renewcommand{\d}{\mathrm{d}}

\newcommand{\C}{\mathbb{C}}

\usepackage{amssymb}

\newcounter{bla}

\journal{Computer Physics Communications}

\begin{document}

\begin{frontmatter}



\title{QEDtool: A Python package for numerical quantum information in quantum electrodynamics}


\author[a]{Jesse Smeets\corref{author}}
\author[b]{Preslav Asenov}
\author[b]{Alessio Serafini}

\cortext[author] {Corresponding author.\\\textit{E-mail address:} j.smeets.physics@gmail.com}
\address[a]{Department of Applied Physics and Science Education, Eindhoven University of Technology, P.O. Box 513, 5600 MB Eindhoven, The Netherlands}
\address[b]{Department of Physics and Astronomy, University College London, \\ Gower Street, London WC1E 6BT, United Kingdom}

\begin{abstract}
This is the manual of the first version of \texttt{QEDtool}, an object-oriented Python package that performs numerical quantum electrodynamics calculations, with focus on full state reconstruction in the internal degrees of freedom, correlations and entanglement quantification. Our package rests on the evaluation of Feynman amplitudes in the momentum-helicity basis within a relativistic framework. Users can specify both pure and mixed initial scattering states in polarization space. From the specified initial state and Feynman amplitudes, \texttt{QEDtool} reconstructs correlations that fully characterize the quantum polarization and entanglement within the final state. These quantities can be expressed in any inertial frame by arbitrary, built-in Lorentz transformations. 







\noindent \textbf{PROGRAM SUMMARY/NEW VERSION PROGRAM SUMMARY}

\begin{small}
\noindent
{\em Program Title:} QEDtool                                          \\
{\em Developer's repository link:} https://github.com/jsmeets2k/qedtool \\
{\em Licensing provisions:} MIT \\
{\em Programming language:} Python                                   \\
{\em Nature of problem:}\\
  Technologies such as quantum-entangled positron emission tomography [1], quantum lithography [2], and quantum free electron lasers [3] are subjected to quantum mechanical processes at relativistic energy scales, best described by quantum field theory (QFT). The dominant interactions probed by the aforementioned technologies are electromagnetic [with quantum electrodynamics (QED) as the corresponding QFT]. Tree-level QED processes form the leading contributions to these dynamics. The analytical determination of tree-level cross sections, entanglement, and spin correlations for arbitrary initial multi-particle states for these processes (and Lorentz transformations thereon) is challenging and time-consuming. \\
{\em Solution method:}\\
  QEDtool provides a basis in which users can easily compute Feynman amplitudes, cross sections, entanglement correlations and spin correlations numerically. Users can define 4-momenta and particles in which polarizations and Green's functions are automatically calculated in the momentum-helicity basis. Additionally, momentum-helicity eigenstates and (classical and quantum) superpositions thereof can be formed, which define the initial state. Built-in functions construct the momentum-projected final state using Feynman amplitudes and the initial state. This allows one to study spin-behavior in Lorentz transformations, and how it impacts the aforementioned correlations and their spatial distributions. \\
{\em Additional comments including restrictions and unusual features:}\\
  All quantum states and their properties are considered in the polarization-helicity Hilbert space only, where a von Neumann measurement is applied in the momentum Hilbert space (as opposed to a partial trace of momentum degrees of freedom); no momentum-space orthogonality and entanglement are taken into account. The particles' 4-momenta are used to determine the Lorentz transformation properties of the momentum-helicity eigenstates.
   \\

{\footnotesize
\begin{center}
© 2026. This manuscript version is made available under the CC-BY-NC-ND 4.0 license \href{https://creativecommons.org/licenses/by-nc-nd/4.0/}{https://creativecommons.org/licenses/by-nc-nd/4.0/}.
\end{center}}

\end{small}
   \end{abstract}
\end{frontmatter}


\section{Introduction}

The past three decades have witnessed intense research and development in the entire area of quantum technologies, ranging from quantum cryptography~\cite{QuantumCryptography,pirandola19} to quantum sensing and metrology~\cite{QuantumSensing,LIGO} and, of course, quantum computing~\cite{nielsechuang,QuantumComputing}. Several of these endeavors require or make direct use of quantum correlations, a type of correlation which cannot be achieved classically, as it can exhibit a variety of strongly non-local, characteristically quantum, properties~\cite{Nonlocality}. 
This non-locality is a consequence of the tensor product structure of composite quantum systems as well as the superposition principle, which lead to entangled quantum states, a key ingredient in many technological and fundamental applications of quantum mechanics.

Albeit the traditional domain of quantum technologies lies in the, highly controllable, ultra-cold and low-energy regime, as is the case, for instance, for superconducting circuits and trapped particles~\cite{QuantumComputing, QuantumSensing}, recent developments of practical interest have moved to the high-energy tail of the spectrum, as in quantum lithography and in the first preliminary steps towards positron emission tomography (QE-PET)~\cite{QuantumLithography, QE-PET}. The latter is based on the strong entanglement between gamma photons from electron-positron pair annihilation~\cite{AnnihilationPhotons}, which might enable one to filter the annihilation signal from the unwanted background~\cite{QE-PET}. Before the photon pair is detected though, it might experience various scattering events, the leading-order process being Compton scattering. The effects on the entanglement of the photon pair for several Compton scattering processes has been investigated, though a fully comprehensive theoretical framework is still lacking~\cite{PryceWard1947, Caradonna2024, Bordes2024}. A full understanding of such entanglement-led processes is not only interesting from a technological perspective, but also in terms of fundamental research, as the measurement of entangled photons in the MeV regime is very different from the polarization entanglement of optical photons~\cite{Bordes2024}. The entanglement of (optical) photons is exploited in other imaging fields as well, see Refs.~\cite{optical_entanglement, entangled_imaging, entangled_photon_ghost_imaging}.

The development of techniques such as QE-PET requires precise angle and energy-resolved quantum state descriptions. Quantum field theory (QFT) does of course provide a relativistic framework for (high-energy) quantum scattering calculations, allowing one to recover entanglement and polarization correlations~\cite{AnnihilationPhotons, Lotstedt2009, FedidaSerafini2023, deVos2024,blasone1,blasone2,blasone3}. This opens up the possibility for detailed characterizations of scattered quantum states. Since the dominant interaction in most technologies is the electromagnetic interaction, quantum electrodynamics (QED) offers the fundamental description for such high-energy studies of entanglement and quantum correlations.

In view of the current interest in accessing fully quantum coherent features in QED, we introduce here a novel Python-based package (\texttt{QEDtool}) that is capable of numerically calculating tree-level QED scattering amplitudes. Whilst most historical applications of QED are to accurately calculate scattering cross sections and fundamental quantities such as electric and magnetic moments~\cite{Schwinger,g2muon}, \texttt{QEDtool} focuses on the reconstruction of the full quantum state in the helicity basis. By defining the parameters of the initial quantum state (either pure or mixed), users can promptly calculate two-particle helicity or polarization entanglement and $n$-particle correlations of the corresponding post-scattering state. Furthermore, at a lower level, QED offers various building blocks to reconstruct QED scattering amplitudes with full polarization dependence, as well as to switch between different reference frames and coordinate systems. Indeed, the package is conceived hierarchically, with lower-level, detailed customizable primitives feeding into higher-level, pre-packaged functions.

\section{Quantum electrodynamics and quantum information}

All theoretical preliminaries are set out in this section, without assuming any previous knowledge of QED or quantum information. We will be very detailed in this regard, recalling standard information, so as to dispel any ambiguity of notation or convention.
Section~\ref{subsec:units} outlines the units and conventions used within the QED framework. Section~\ref{subsec:S-matrix_rho} explains how the post-scattering state can be constructed from the pre-scattering state and Feynman amplitudes. The Feynman amplitudes are obtained from the perturbative $S$-matrix formalism. We specifically seek the polarized amplitudes, i.e. the amplitudes of momentum-helicity eigenstate scattering. These are calculated from within the momentum-helicity basis, which is presented in Section~\ref{subsec:momentum-helicity_basis}. In Section~\ref{subsec:Lorentz_transformation}, the different representations of Lorentz transformations to carry evaluations over to different frames are discussed. Finally, Section~\ref{subsec:concurrence_Stokes} contains the necessary theory about the degree of entanglement and spin correlations.

\subsection{Units and conventions}\label{subsec:units}

\texttt{QEDtool} works with natural units, i.e. $\hbar = c = \epsilon_0 = 1$ with $\hbar$ the reduced Planck constant, $c$ the speed of light and $\epsilon_0$ the vacuum permittivity. In SI units, these have values $\hbar \approx 1.05\times10^{-34}~\mathrm{J\,s}$, $c \approx 3\times10^8~\mathrm{m\,s^{-1}}$ and $\epsilon_0 \approx 8.85\times10^{-12}~\mathrm{F\,m^{-1}}$. Moreover, $\alpha = e^2/(4\pi\epsilon_0\hbar c) \approx 1/137$ is the (dimensionless) fine structure constant, with $e$ the elementary charge ($e \approx 1.6\times10^{-19}~\mathrm{C}$). Consequentially, momenta, frequencies, energies and masses are all expressed in units of energy. Users can specify the energy units by specifying the order of magnitude in ${\rm eV}$. 

For massive particles with rest mass $m$ and a relativistic 3-momentum $\mathbf{p}$, the relativistic energy-momentum relation reads $\varepsilon_\mathbf{p}^2 = \mathbf{p}^2 + m^2$. Generally, 4-momenta are expressed as $p = (\varepsilon_\mathbf{p}, \mathbf{p}) = \gamma_{\boldsymbol{\beta}} m(1,\boldsymbol{\beta})$, where $\smash{\gamma_{\boldsymbol{\beta}} = (1-\boldsymbol{\beta}^2)^{-1/2}}$ is the Lorentz factor and $\boldsymbol{\beta}$ denotes the 3-velocity. For the flat space-time metric, the mostly minus convention $(+,-,-,-)$ is employed. Therefore, the Lorentzian product of 4-vectors $a$ and $b$ equals $a\cdot b = a^0 b^0 - \mathbf{a} \cdot \mathbf{b}$, where $\mathbf{a}$ and $\mathbf{b}$ are 3-vectors and $\mathbf{a}\cdot\mathbf{b}$ is their Euclidean inner product. In a similar fashion, contractions with the gamma matrices become $\slashed{a} = \gamma\cdot a$, where the gamma matrices are expressed in the chiral basis,
\begin{equation*}
    \gamma^\mu = \begin{pmatrix}
        0 & \sigma^\mu \\
        \Bar{\sigma}^\mu & 0
    \end{pmatrix}\,,
\end{equation*}
with $\sigma^\mu \equiv (\sigma^0, \boldsymbol{\sigma})$ and $\Bar{\sigma}^\mu \equiv (\sigma^0, -\boldsymbol{\sigma})$. Here $\sigma^0$ is the $2\times2$ identity matrix, also referred to as the zeroth Pauli matrix, and $\boldsymbol{\sigma} \equiv (\sigma^1, \sigma^2, \sigma^3)$ are the $x$, $y$ and $z$ Pauli matrices.

\texttt{QEDtool} expresses vector components in the standard coordinate systems: (1) Cartesian coordinates $(x,y,z)$; (2) Cylindrical coordinates $(\rho,\phi,z)$, where $\rho = \sqrt{x^2+y^2}$ and $\phi$ is the angle with respect to the $x$-axis; (3) Spherical coordinates $(r,\theta,\phi)$ with $\theta$ as the polar angle, i.e. the angle between $\mathbf{v}$ and the $z$-axis. Angle $\phi$ is the angle between $\mathbf{v}$ and the $x$-axis, referred to as the azimuthal angle.

We make use of common orthonormal photon and fermion polarization bases; left-right $\{\mathrm{L}, \mathrm{R}\}$, horizontal-vertical $\{\mathrm{H}, \mathrm{V}\}$, and diagonal-antidiagonal $\{\mathrm{D}, \mathrm{A}\}$. The basis of choice for \texttt{QEDtool} is the $\{\mathrm{L},\mathrm{R}\}$ basis. All other aforementioned polarizations can be expressed in terms of L- and R-polarization as
\begin{equation}\label{eq:polarization_conventions}
    \begin{aligned}
        &|\mathrm{H}\rangle = \frac{1}{\sqrt{2}}\big( |\mathrm{L}\rangle + |\mathrm{R}\rangle \big)\,, \qquad\quad &&|\mathrm{V}\rangle = \frac{\i}{\sqrt{2}}\big( |\mathrm{L}\rangle - |\mathrm{R}\rangle \big)\,, \\[0.2cm]
        &|\mathrm{D}\rangle = \frac{1}{2}\big[ (1+\i)|\mathrm{L}\rangle + (1-\i)|\mathrm{R}\rangle \big]\,, \qquad\quad 
        &&|\mathrm{A}\rangle = \frac{1}{2}\big[ (1-\i)|\mathrm{L}\rangle + (1+\i)|\mathrm{R}\rangle \big]\,.
    \end{aligned}
\end{equation}
Note that $|\mathrm{L}\rangle$ and $|\mathrm{R}\rangle$, i.e.~circular polarizations, are helicity eigenstates with eigenvalues $h = \pm1$ for photons and $h = \pm1/2$ for electrons. Throughout this manual, we will refer to a particle's ``handedness'' instead of its helicity.\footnote{The helicity normalized to unity; $\mathrm{sgn}\,h$.}

\subsection{Density operator formalism}\label{subsec:S-matrix_rho}

Consider an $n$-particle quantum state $|\psi(t)\rangle$ and two limits thereof; $\smash{|\psi^{\text{(in)}}\rangle = |\psi(t\to-\infty)\rangle}$ and $\smash{|\psi^{\text{(out)}}\rangle = |\psi(t\to\infty)\rangle}$, referred to as the in and out-asymptotes. They are free states due to their infinite separation. The in- and out-asymptotes are related through the scattering operator $S$, which contains the interacting Hamiltonian. This relation reads 
\begin{equation}\label{eq:S-matrix_pure}
    |\psi^{\text{(out)}}\rangle = S|\psi^{\text{(in)}}\rangle\,.
\end{equation}
Users of \texttt{QEDtool} specify the (generally mixed) in-asymptote. The probability amplitude to transition from the in-asymptote to some state $|\phi\rangle$, is given by $\mathcal{S} \equiv \langle\phi|S|\psi^{\text{(in)}}\rangle$. Here, $|\phi\rangle$ is some other state also evaluated at $t\to+\infty$. Since \texttt{QEDtool} performs quantum scattering calculations with general in-asymptotes that can also be mixed in polarization space, quantum states are represented by their density operators $\rho$. Consequentially, Eq.~(\ref{eq:S-matrix_pure}) then becomes
\begin{equation}\label{eq:S-matrix_mixed}
    \rho^{\text{(out)}} = S\rho^{\text{(in)}}S^\dagger\,.
\end{equation}
We define a general pure $n$-particle in-state as a collection of states
\begin{equation}\label{eq:general_in-state}
    |\Phi_j, j\rangle = \sum_\alpha \int\frac{\d^{3n}\mathsf{p}}{(2\pi)^{3n}}\frac{1}{\prod_{k=1}^n\sqrt{2\varepsilon_k}}\,c_{j\alpha}\Phi_j(\mathsf{p})\,|\mathsf{p},\alpha\rangle\,,
\end{equation}
with $\varepsilon_k = \sqrt{|\mathbf{p}_k|^2 + m_k^2}$ where $m_k$ is the mass of the $k$th particle. In Eq.~(\ref{eq:general_in-state}), $\mathsf{p} \equiv \{p_1,...,p_n\}$ denotes the set of initial 4-momenta and $|j\rangle$ is a superposition of helicity eigenstates $|\alpha\rangle$ with coefficients $c_{j\alpha}\in\C$ and $\alpha\in\{\mathrm{L},\mathrm{R}\}^{\otimes n}$. For the integral we adopt the notation $\d^{3n}\mathsf{p} \equiv \d^3p_1 \cdots \d^3p_n$, and $\Phi_j(\mathsf{p})$ signifies the $n$-particle momentum wave function, which is normalized as
\begin{equation*}
    \int\frac{\d^{3n}\mathsf{p}}{(2\pi)^{3n}}\,|\Phi_j(\mathsf{p})|^2 = 1\,.
\end{equation*}
Let us now assume that the in-asymptote is actually a mixed state given by the convex combination of the pure states
\begin{equation}\label{eq:convex}
    \rho^{\text{(in)}} = \sum_j w_j |\Phi_j, j\rangle\langle\Phi_j, j|
\end{equation}
weighted by classical probabilities $w_j$ such that $\sum_j w_j = 1$. The density operator that corresponds to Eq.~(\ref{eq:general_in-state}), taking the classical probabilities into account, becomes
\begin{equation}\label{eq:rho_in}
    \rho^{\text{(in)}} = \sum_j \sum_{\alpha,\beta} \int_\mathsf{p}\int_{\tilde{\mathsf{p}}}\,w_jc_{j\alpha}c^*_{j\beta}\Phi_j(\mathsf{p})\,\Phi^*_j(\tilde{\mathsf{p}})\,|\mathsf{p},\alpha\rangle\langle\tilde{\mathsf{p}},\beta|\,,
\end{equation}
where we introduced the notation
\begin{equation*}
    \int_\mathsf{p} \equiv \int\frac{\d^{3n}\mathsf{p}}{(2\pi)^{3n}}\frac{1}{\prod_{k=1}^n\sqrt{2\varepsilon_k}}\,.
\end{equation*}
We study scattering to momentum eigenstates. However, $\rho^{\text{(out)}}$ from Eq.~(\ref{eq:S-matrix_mixed}) is not necessarily a momentum eigenstate; it is the out-asymptote that corresponds to Eq.~(\ref{eq:rho_in}). We therefore consider the ideal filtering of outgoing momenta, by applying the projection $\Pi_{\Bar{\mathsf{p}}} \equiv \sum_{\bar{\alpha}}|\Bar{\mathsf{p}},\bar{\alpha}\rangle\langle\Bar{\mathsf{p}},\bar{\alpha}|$, which projects $\smash{\rho^{\text{(out)}}}$ onto a momentum-helicity eigenstate with 4-momenta $\Bar{\mathsf{p}}$, keeping the helicities intact. Thus, we obtain
\begin{equation}\label{eq:projected_out}
\begin{aligned}
    \Pi_{\bar{\mathsf{p}}}\,S\rho^{\text{(in)}}S^\dagger\,\Pi_{\bar{\mathsf{p}}} &= \sum_{\alpha,\beta} \sum_{\bar{\alpha},\bar{\beta}} \sum_j \int_\mathsf{p}\int_{\tilde{\mathsf{p}}}\,\,w_jc_{j\alpha}c^*_{j\beta}\,\Phi_j(\mathsf{p})\,\Phi^*_j(\tilde{\mathsf{p}}) \\[-0.2cm]
    &\hspace{3.3cm}\times\mathcal{S}_{\bar{\alpha}\alpha}(\mathsf{p}\to\bar{\mathsf{p}})\,\mathcal{S}^*_{\bar{\beta}\beta}(\tilde{\mathsf{p}}\to\bar{\mathsf{p}})\,|\bar{\mathsf{p}},\bar{\alpha}\rangle\langle\bar{\mathsf{p}},\bar{\beta}|,
\end{aligned}
\end{equation}
with $\mathcal{S}_{\bar{\alpha}\alpha}(\mathsf{p}\to\bar{\mathsf{p}}) \equiv \langle\bar{\mathsf{p}},\bar{\alpha}|S|\mathsf{p},\alpha\rangle$ being the $S$-matrix elements. From these $S$-matrix elements, we define the Feynman amplitudes $\i\mathcal{M}_{\bar{\alpha}\alpha}$ as
\begin{equation*}
    \mathcal{S}_{\bar{\alpha}\alpha}(\mathsf{p}\to\bar{\mathsf{p}}) = \i\mathcal{M}_{\bar{\alpha}\alpha}(\mathsf{p}\to\bar{\mathsf{p}}) \times (2\pi)^4\delta^4\big[\textstyle\sum_{k=1}^n (\bar{p}_k - p_k)\big]\,.
\end{equation*}
With this definition, Eq.~(\ref{eq:projected_out}) becomes
\begin{equation}\label{eq:projected_out-M}
\begin{aligned}
    \Pi_{\bar{\mathsf{p}}}\,S\rho^{\text{(in)}}S^\dagger\,\Pi_{\bar{\mathsf{p}}} &= (2\pi)^8\sum_{\alpha,\beta} \sum_{\bar{\alpha},\bar{\beta}} \sum_j \int_\mathsf{p}\int_{\tilde{\mathsf{p}}}\,\,w_jc_{j\alpha}c^*_{j\beta}\,\Phi_j(\mathsf{p})\,\Phi^*_j(\tilde{\mathsf{p}}) \\[-0.1cm]
    &\hspace{4.4cm}\times\delta^4\big[\textstyle\sum_{k=1}^n (\bar{p}_k - p_k)\big]\,\delta^4\big[\textstyle\sum_{k=1}^n (\bar{p}_k - \tilde{p}_k)\big] \\[0.25cm]
    &\hspace{4.4cm}\times\mathcal{M}_{\bar{\alpha}\alpha}(\mathsf{p}\to\bar{\mathsf{p}})\,\mathcal{M}^*_{\bar{\beta}\beta}(\tilde{\mathsf{p}}\to\bar{\mathsf{p}})\,|\bar{\mathsf{p}},\bar{\alpha}\rangle\langle\bar{\mathsf{p}},\bar{\beta}|\,.
\end{aligned}
\end{equation}
To normalize the projected out-state, we normalize Eq.~(\ref{eq:projected_out-M}) by its trace,
\begin{equation*}
    \rho^{\text{(out)}}_{\bar{\mathsf{p}}} \equiv \frac{\Pi_{\bar{\mathsf{p}}}\,S\rho^{\text{(in)}}S^\dagger\,\Pi_{\bar{\mathsf{p}}}}{\mathrm{tr}\big(\Pi_{\bar{\mathsf{p}}}\,S\rho^{\text{(in)}}S^\dagger\,\Pi_{\bar{\mathsf{p}}}\big)}\,.
\end{equation*}
This becomes
\begin{equation}\label{eq:projected-out}
\begin{aligned}
    \rho^{\text{(out)}}_{\bar{\mathsf{p}}}
    &= \frac{1}{\langle \bar{\mathsf{p}}, \bar{\alpha} | \bar{\mathsf{p}},\bar{\alpha} \rangle}\frac{1}{\partial_{\bar{\mathsf{p}}}W}\sum_{\alpha,\beta} \sum_{\bar{\alpha},\bar{\beta}} \sum_j \int_\mathsf{p}\int_{\tilde{\mathsf{p}}}\,\,w_jc_{j\alpha}c^*_{j\beta}\,\Phi_j(\mathsf{p})\,\Phi^*_j(\tilde{\mathsf{p}}) \\[-0.1cm]
    &\hspace{5.7cm}\times\delta^4\big[\textstyle\sum_{k=1}^n (\bar{p}_k - p_k)\big]\,\delta^4\big[\textstyle\sum_{k=1}^n (\bar{p}_k - \tilde{p}_k)\big] \\[0.25cm]
    &\hspace{5.7cm}\times\mathcal{M}_{\bar{\alpha}\alpha}(\mathsf{p}\to\bar{\mathsf{p}})\,\mathcal{M}^*_{\bar{\beta}\beta}(\tilde{\mathsf{p}}\to\bar{\mathsf{p}})\,|\bar{\mathsf{p}},\bar{\alpha}\rangle\langle\bar{\mathsf{p}},\bar{\beta}|\,.
\end{aligned}
\end{equation}
Here we defined
\begin{equation}\label{eq:dW}
\begin{aligned}
    \frac{\partial W}{\partial\bar{\mathsf{p}}} &\equiv \frac{(2\pi)^8}{\langle \bar{\mathsf{p}}, \bar{\alpha}|\bar{\mathsf{p}},\bar{\alpha}\rangle}\sum_\xi \sum_{\alpha,\beta} \sum_j \int_\mathsf{p}\int_{\tilde{\mathsf{p}}}\,w_jc_{j\alpha}c^*_{j\beta}\,\Phi_j(\mathsf{p})\,\Phi^*_j(\tilde{\mathsf{p}}) \\
    &\hspace{5cm}\times\delta^4\big[\textstyle\sum_{k=1}^n (\bar{p}_k - p_k)\big]\,\delta^4\big[\textstyle\sum_{k=1}^n (\bar{p}_k - \tilde{p}_k)\big] \\[0.25cm]
    &\hspace{5cm}\times\mathcal{M}_{\xi\alpha}(\mathsf{p}\to\bar{\mathsf{p}})\,\mathcal{M}^*_{\xi\beta}(\tilde{\mathsf{p}}\to\bar{\mathsf{p}})\,,
\end{aligned}
\end{equation}
where $\partial_{\bar{\mathsf{p}}} \equiv \prod_k^n\partial_{\varepsilon_k}\partial_{|\mathbf{p}_k|}\partial_{\Omega_k}$ and $\varepsilon_{k}$, $|\mathbf{p}_k|$ and $\Omega_k$ are respectively the energy, Euclidean 3-momentum norm and solid angle of the $k$th particle. \texttt{QEDtool} calculates the aforementioned quantities for initial states that are momentum eigenstates. In that case $\Phi_j(\mathsf{p}) = \delta^{4n}(\mathsf{p}-\mathsf{p}_0)$ for the definite 4-momenta $\mathsf{p}_0$. Performing the integrals in Eqs.~(\ref{eq:projected-out}, \ref{eq:dW}) then removes the integral signs and $\mathsf{p},\tilde{\mathsf{p}} \to \mathsf{p}_0$.

Another insightful quantity is the probability for a specific polarization state to exit the scattering event. For this, we project $\rho^{\text{(out)}}_{\bar{\mathsf{p}}}$ onto the density operator $\varrho$ of the sought quantum state. The angular probability of finding state $\varrho$ is then given by $\smash{\partial_{\bar{\mathsf{p}}} W_\varrho \equiv \mathrm{tr}\big( \rho^{\text{(out)}}_{\bar{\mathsf{p}}} \varrho \big)}$.

Eqs.~(\ref{eq:projected-out}, \ref{eq:dW}) contain Dirac deltas, e.g. $\delta^4(p-q) = \delta(\varepsilon_\mathbf{p}-\varepsilon_\mathbf{q})\,\delta^3(\mathbf{p}-\mathbf{q})$, which can be regulated by considering the quantization volume $\mathcal{V}$ and interaction time $\mathcal{T}$, as
\begin{equation*}
    \delta^3(\mathbf{p}-\mathbf{q}) = \frac{\mathcal{V}}{(2\pi)^3}\delta_{\mathbf{p},\mathbf{q}}\,, \qquad\qquad \delta(\varepsilon_\mathbf{p}-\varepsilon_\mathbf{q}) = \frac{1}{2\pi}\int_{-\mathcal{T}/2}^{\mathcal{T}/2}\d t\,\exp[\i(\varepsilon_\mathbf{p}-\varepsilon_\mathbf{q})t]\,,
\end{equation*}
which imply $\mathcal{V} \equiv (2\pi)^3\delta^3(\mathbf{0})$ and $\mathcal{T}\equiv2\pi\delta(0)$. As a consequence, single-particle momentum eigenstates are normalized as $\langle p|p\rangle = 2\mathcal{V}\varepsilon_\mathbf{p}$. For the scattering of momentum eigenstates, Eqs.~(\ref{eq:projected-out}, \ref{eq:dW}) will contain squared Dirac deltas, which may be recast as~\cite{Weinberg}
\begin{equation*}
    \big[\delta^4(p-q)\big]^2 = \frac{\mathcal{TV}}{(2\pi)^4}\delta^4(p-q)\,.
\end{equation*}
Focusing on 2-to-2 particle ($n=2$) momentum eigenstate scattering, Eq.~(\ref{eq:dW}) becomes
\begin{equation*}
\begin{aligned}
    \frac{\partial W}{\partial\bar{\mathsf{p}}} &= (2\pi)^4\mathcal{TV}\sum_\xi \sum_{\alpha,\beta} \sum_j \frac{1}{4\varepsilon_1\varepsilon_2}\,\frac{1}{4\mathcal{V}^2 \bar{\varepsilon}_1\bar{\varepsilon}_2}\,w_jc_{j\alpha}c^*_{j\beta} \\
    &\hspace{3.6cm}\times\delta^4\big( \bar{p}_1 + \bar{p}_2 - p_1 - p_2 \big) \\[0.25cm]
    &\hspace{3.6cm}\times\mathcal{M}_{\xi\alpha}(\mathsf{p}\to\bar{\mathsf{p}})\,\mathcal{M}^*_{\xi\beta}(\mathsf{p}\to\bar{\mathsf{p}})\,,
\end{aligned}
\end{equation*}
Defining the two-body relativistically invariant phase space differential~\cite{PeskinSchroeder},
\begin{equation*}
    \d\Pi = (2\pi)^4 \delta^4\big( \bar{p}_1 + \bar{p}_2 - p_1 - p_2 \big)\,\frac{\d^3\bar{p}_1}{(2\pi)^3}\frac{1}{2\bar{\varepsilon}_1}\,\frac{\d^3\bar{p}_2}{(2\pi)^3}\frac{1}{2\bar{\varepsilon}_2}\,,
\end{equation*}
we obtain
\begin{align}\label{eq:dW/dPi}
    \frac{\partial W}{\partial\Pi} &= \frac{\mathcal{T}}{\mathcal{V}}\sum_\xi \sum_{\alpha,\beta} \sum_j \frac{1}{4\varepsilon_1\varepsilon_2}\,w_jc_{j\alpha}c^*_{j\beta} \,\mathcal{M}_{\xi\alpha}(\mathsf{p}\to\bar{\mathsf{p}})\,\mathcal{M}^*_{\xi\beta}(\mathsf{p}\to\bar{\mathsf{p}}) \nonumber\\
    &\equiv \frac{\mathcal{T}}{\mathcal{V}}\frac{1}{4\varepsilon_1\varepsilon_2}\frac{\partial\mathcal{P}}{\partial\Pi}\,.
\end{align}
The differential probabilities are not directly measurable, however a quantity that is often calculated in standard QFT literature is $\partial_\Pi\mathcal{P}$, as defined in Eq.~(\ref{eq:dW/dPi}). For an initial unpolarized $n$-particle state, i.e. $\smash{\rho_{\alpha\beta} = 2^{-n}\delta_{\alpha\beta}}$, we have $\smash{\partial_\Pi\mathcal{P} = 2^{-n}\sum_{\text{spins}}|\mathcal{M}|^2}$. The differential cross section is defined as~\cite{Schwartz}
\begin{equation*}
    \d\sigma = \frac{\mathcal{V}}{\mathcal{T}}\frac{1}{|\boldsymbol{\beta}_1 - \boldsymbol{\beta}_2|}\d W\,,
\end{equation*}
with $\boldsymbol{\beta}_{1(2)}$ being the initial 3-velocities of particle 1(2), which leads to the expression
\begin{equation*}
\begin{aligned}
    \frac{\partial\sigma}{\partial\Pi} &= \sum_\xi \sum_{\alpha,\beta} \sum_j \frac{1}{4\varepsilon_1\varepsilon_2}\frac{1}{|\boldsymbol{\beta}_1 - \boldsymbol{\beta}_2|}\,w_jc_{j\alpha}c^*_{j\beta} \,\mathcal{M}_{\xi\alpha}(\mathsf{p}\to\bar{\mathsf{p}})\,\mathcal{M}^*_{\xi\beta}(\mathsf{p}\to\bar{\mathsf{p}})\,,
\end{aligned}
\end{equation*}
which is a measurable quantity.

\subsection{S-matrix formalism in the momentum-helicity basis}\label{subsec:momentum-helicity_basis}

\texttt{QEDtool} works with the QED Lagrangian density operator
\begin{equation}\label{eq:Lagrangian}
    \mathcal{L} = \bar{\psi}(\i\slashed{\partial} - m)\psi - \frac{1}{4}F_{\mu\nu}F^{\mu\nu} - e\bar{\psi}\slashed{A}\psi\,.
\end{equation}
with $\psi(x)$ the Dirac field and $A(x)$ is the photon field. $F_{\mu\nu}(x) \equiv \partial_\mu A_\nu - \partial_\nu A_\mu$ is the electromagnetic field tensor. In Eq.~(\ref{eq:Lagrangian}) $\smash{\bar{\psi}(x) \equiv \psi^\dagger(x)\gamma^0}$ denotes the Dirac adjoint of the Dirac field. By transforming to the interaction picture and using $\mathcal{H}_\mathrm{I} = - \mathcal{L}_\mathrm{I} = e\bar{\psi}\slashed{A}\psi$, the $S$-matrix elements from Eq.~(\ref{eq:projected_out}) are~\cite{PeskinSchroeder}
\begin{equation}\label{eq:S_H-int}
    \langle\bar{\mathsf{p}},\bar{\alpha}|S|\mathsf{p},\alpha\rangle = \langle\bar{\mathsf{p}},\bar{\alpha}|\,T\Bigg\{ \exp\!\Bigg[-\i\int\d^4x\,\mathcal{H}_{\text{I}}(x)\Bigg] \Bigg\}\,|\mathsf{p},\alpha\rangle
\end{equation}
where $T$ denotes the time-ordering operator. As customary, we shall be considering these $S$-matrix elements to the lowest perturbative order, which is reliable due to the smallness of the coupling constant $\alpha$.

For the field operators, \texttt{QEDtool} utilizes the momentum-helicity basis. In the Lorenz gauge $\partial \cdot A = 0$ and in the absence of currents, the electromagnetic field satisfies the vacuum Maxwell equations $\partial^2 A = 0$, whose quantized solution expanded in Fourier modes is~\cite{PeskinSchroeder, Schwartz}
\begin{equation}\label{eq:photon_field}
    A(x) = \sum_\lambda \int \frac{\d^3p}{(2\pi)^3}\frac{1}{\sqrt{2|\mathbf{p}|}}\Big[ c_{\lambda,\mathbf{p}}\epsilon_\lambda(p)\exp(-\i p\cdot x) + c^{\dagger}_{\lambda,\mathbf{p}}\epsilon_\lambda^*(p)\exp(\i p\cdot x) \Big]\,,
\end{equation}
where $\smash{c^{(\dagger)}_{\lambda,\mathbf{p}}}$ are the momentum space field operators that annihilate(create) a photon with 4-momentum $p = (|\mathbf{p}|,\mathbf{p})$ and polarization mode $\lambda$, and $\epsilon_\lambda(p)$ is the photon's 4-polarization. The momentum space field operators satisfy the commutation relations
\begin{equation}\label{eq:commutation}
    \Big[ c_{\lambda,\mathbf{p}}, c^\dagger_{\lambda'\!,\mathbf{p}'} \Big] = \delta_{\lambda\lambda'} \delta^3(\mathbf{p}-\mathbf{p}')\,, \qquad\quad \Big[ c_{\lambda,\mathbf{p}}, c_{\lambda'\!,\mathbf{p}'} \Big] = \Big[ c^\dagger_{\lambda,\mathbf{p}}, c^\dagger_{\lambda'\!,\mathbf{p}'} \Big] = 0\,.
\end{equation}
The polarization basis is $\lambda \in \{\mathrm{L},\mathrm{R},\mathrm{F},\mathrm{B}\}$, where L(R) stands for left(right)-handed and F(B) is forward(backward). For a photon with definite 4-momentum
\begin{equation}\label{eq:4-momentum}
    p = (\varepsilon_\mathbf{p},\,|\mathbf{p}|\sin\theta\cos\phi,\,|\mathbf{p}|\sin\theta\sin\phi,\,|\mathbf{p}|\cos\theta)\,,
\end{equation}
(with $\varepsilon_\mathbf{p} = |\mathbf{p}|$ such that $p$ is light-like) the two transverse 4-polarizations are given by
\begin{equation*}
    \epsilon_{\mathrm{L}(\mathrm{R})}(p) = \dfrac{1}{\sqrt{2}}(
        0,
        \cos\theta\cos\phi \pm \i\sin\phi,
        \cos\theta\sin\phi \mp \i\cos\phi,
        -\sin\theta
    )\,,
\end{equation*}
with $p\cdot\epsilon_{\mathrm{L}(\mathrm{R})}(p) = 0$. The forward and backward 4-polarizations are $\epsilon_{\mathrm{F}(\mathrm{B})}(p) = (1,\pm\hat{\mathbf{p}})$~\cite{PeskinSchroeder}. Due to the gauge symmetry, external photons are always transverse, hence the forward and backward polarizations are nonphysical. Off-shell photons contain longitudinal polarization components, e.g. photons responsible for the Coulomb interaction are purely longitudinal.

The Dirac field entails both electrons and positrons, and it satisfies the Dirac equation $(\i\slashed{\partial} - m)\psi = 0$. The corresponding solution in second quantization, expanded into Fourier modes equals~\cite{PeskinSchroeder, Schwartz}
\begin{equation*}
    \psi(x) = \sum_\sigma \int\frac{\d^3p}{(2\pi)^3}\frac{1}{\sqrt{2\varepsilon_\mathbf{p}}}\,\Big[ a_{\sigma,\mathbf{p}}u_\sigma(p)\exp(-\i p\cdot x) + b^{\dagger}_{\sigma,\mathbf{p}}v_\sigma(p)\exp(\i p\cdot x) \Big]\,,
\end{equation*}
where $a^{(\dagger)}_{\sigma,\mathbf{p}}$ and $b^{(\dagger)}_{\sigma,\mathbf{p}}$ annihilate(create) electrons and positrons respectively, with handedness $\sigma$ and 3-momentum $\mathbf{p}$. These operators anticommute;
\begin{equation}\label{eq:anticommutation}
    \Big\{ a_{\sigma,\mathbf{p}}, a^\dagger_{\sigma'\!,\mathbf{p}'} \Big\} = \Big\{ b_{\sigma,\mathbf{p}}, b^\dagger_{\sigma'\!,\mathbf{p}'} \Big\} = (2\pi)^3\delta_{\sigma\sigma'} \delta^3(\mathbf{p}-\mathbf{p}')\,,
\end{equation}
and all other anticommutators are zero. Moreover, $u_\sigma(p)$ and $v_\sigma(p)$ are Dirac spinors that satisfy the momentum space Dirac equations $(\slashed{p} - m)u_\sigma(p) = 0$ and $(\slashed{p} + m)v_\sigma(p) = 0$. For electrons and positrons with the 4-momentum from Eq.~(\ref{eq:4-momentum}), the solutions are of the form
\begin{equation*}
    u_{\sigma}(p) = \begin{pmatrix}
        \sqrt{\varepsilon_\mathbf{p} - \sigma|\mathbf{p}|}\,\chi_{\sigma}(p) \\[0.1cm]
        \sqrt{\varepsilon_\mathbf{p} + \sigma|\mathbf{p}|}\,\chi_{\sigma}(p)
    \end{pmatrix}, \quad\quad
    v_{\sigma}(p) = \begin{pmatrix}
        -\sigma\sqrt{\varepsilon_\mathbf{p} + \sigma|\mathbf{p}|}\,\chi_{-\sigma}(p) \\[0.1cm]
        \sigma\sqrt{\varepsilon_\mathbf{p} - \sigma|\mathbf{p}|}\,\chi_{-\sigma}(p)
    \end{pmatrix},
\end{equation*}
with the two-component helicity eigenspinors~\cite{PeskinSchroeder}
\begin{equation*}
    \chi_{\mathrm{R}}(p) = \begin{pmatrix}
        \cos(\theta/2) \\
        \exp(\i\phi)\sin(\theta/2)
    \end{pmatrix},
    \qquad
    \chi_{\mathrm{L}}(p) = \begin{pmatrix}
        -\exp(-\i\phi)\sin(\theta/2) \\
        \cos(\theta/2)
    \end{pmatrix}.
\end{equation*}
The two polarizations modes are $\sigma\in\{\mathrm{L},\mathrm{R}\}$, where L(R) corresponds to $\sigma = \mp1$.

A central quantity in quantum scattering theory is the Green's function, also referred to as the propagator, which are heuristically interpreted as the amplitude for a field to propagate from $x$ to $x'$ in space-time. For the photon field, this amplitude is given by the two-point correlator $\langle 0|T\{A_\mu(x)A_\nu(x')\}|0\rangle$, where $T$ denotes time-ordering. Evaluating this correlation with the quantized photon field from Eq.~(\ref{eq:photon_field}), one obtains an expression of the form
\begin{align}\label{eq:D_munu}
    D_{\mu\nu}(x,x') &= \int\frac{\d^4q}{(2\pi)^4}\frac{-\i g_{\mu\nu}}{q^2 + \i0^+}\exp[-\i q\cdot(x-x')] \nonumber\\[0.2cm]
    &\equiv \int\frac{\d^4q}{(2\pi)^4}\,\Tilde{D}_{\mu\nu}(q)\exp[-\i q\cdot(x-x')] \,,
\end{align}
where the term $+\i0^+$ with $\smash{0^+ \equiv \lim_{\epsilon\downarrow0} \epsilon}$ ensures the Feynman prescription (time-ordering) and $g_{\mu\nu}$ is the metric tensor. Here $\smash{\Tilde{D}_{\mu\nu}(q)}$ is the momentum space photon propagator, which is also the Green's function of the momentum space Maxwell equations $q^2 \tilde{A}(q) = 0$ with $\tilde{A}(q)$ as the momentum space photon field. The amplitude for an electron or positron to propagate from event $x$ to $x'$ is given by the correlation $\langle 0|T\{\psi(x)\bar{\psi}(x')\}|0\rangle$. Evaluating this correlation gives
\begin{align}\label{eq:G}
    G(x,x') &= \int\frac{\d^4q}{(2\pi)^4}\,\frac{\i(\slashed{q}+m)}{q^2-m^2+\i0^+}\exp[-\i q\cdot(x-x')] \nonumber\\[0.2cm] 
    &\equiv \int\frac{\d^4q}{(2\pi)^4}\,\tilde{G}(q)\exp[-\i q\cdot(x-x')]\,,
\end{align}
which is also the Green's function of the Dirac equation.

Note that the $S$-matrix elements are vacuum expectation values of time-ordered products of field operators. Consider e.g. the electron scattering matrix element
\begin{equation*}
    \langle p'_1,\sigma'_1;p'_2,\sigma'_2|\,S\,|p_1,\sigma_1;p_2,\sigma_2\rangle \propto \langle0|\,a_{\sigma'_1,\mathbf{p}'_1} a_{\sigma'_2,\mathbf{p}'_2}\,S\,a^\dagger_{\sigma_1,\mathbf{p}_1} a^\dagger_{\sigma_2,\mathbf{p}_2}|0\rangle
\end{equation*}
where we defined momentum-helicity eigenstates $|p,\sigma\rangle \equiv \sqrt{2\varepsilon_\mathbf{p}}a^\dagger_{\sigma,\mathbf{p}}|0\rangle$ and $S$ contains time-ordered products of fields [see Eq.~(\ref{eq:S_H-int})]. Wick's theorem allows one to write $S$-matrix elements in terms of the aforementioned Green's functions and polarizations. Namely, a time-ordered operator product equals the normal-ordered product minus all possible products of Wick contracted pairs. For an operator product $AB$, their Wick contraction is written as $A^\bullet B^\bullet$. Two important Wick contractions are $\psi(x)^\bullet\bar{\psi}(x')^\bullet = G(x,x')$ and $\smash{A_\mu(x)^\bullet A_\nu(x')^\bullet = D_{\mu\nu}(x,x')}$. These represent internal fermion and photon lines respectively in Feynman diagrams. The external lines are from $\psi(x)^\bullet a^{\dagger\bullet}_{\sigma,\mathbf{p}}|0\rangle = u_\sigma(p)\exp(-\i p\cdot x)|0\rangle$ for electrons, and $\smash{A(x)^\bullet c^{\dagger\bullet}_{\lambda,\mathbf{p}}|0\rangle = \epsilon_\lambda(p)\exp(-\i p\cdot x)|0\rangle}$ for photons. We refer the reader to Refs.~\cite{PeskinSchroeder,Schwartz,FetterWalecka} for derivations that elucidate the connection between Feynman rules and Wick contractions. \texttt{QEDtool} operates in momentum space, where external and internal lines represent momentum space polarizations and propagators respectively. Below we present the momentum space Feynman rules for QED:
\vspace{0.2cm}
\begin{alignat}{3}
    &\text{Dirac propagator:}\qquad 
    &&\vcenter{\hbox{\includegraphics[width=2cm]{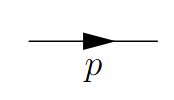}}}
    \quad
    = \frac{-\i (\slashed{p} + m)}{p^2 - m^2 + \i 0^+} \label{eq:Dirac_propagator} \\[0.2cm]
    &\text{Photon propagator:}\qquad
    &&\vcenter{\hbox{\includegraphics[width=2cm]{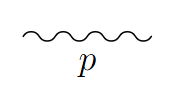}}}
    \quad
    = \frac{-\i g_{\mu\nu}}{p^2 + \i 0^+} \label{eq:photon_propagator} \\[0.2cm]
    &\text{QED vertex:}\qquad
    &&\vcenter{\hbox{\includegraphics[width=2cm]{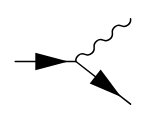}}}
    \quad
    = -\i Q e \gamma^\mu \label{eq:vertex} \\[0.2cm]
    &\text{External fermions:}\qquad
    &&\begin{cases}
         &\vcenter{\hbox{\includegraphics[width=2cm]{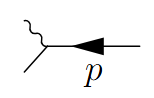}}}
        \quad
        = u_\sigma(p) \qquad \text{(initial)} \\[0.3cm]
         &\vcenter{\hbox{\includegraphics[width=2cm]{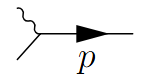}}}
        \quad
        = \bar{u}_\sigma(p) \qquad \text{(final)}
    \end{cases} \label{eq:external_fermions}\\[0.2cm]
    &\text{External antifermions:}\qquad
    &&\begin{cases}
         &\vcenter{\hbox{\includegraphics[width=2cm]{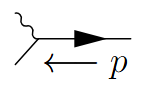}}}
        \quad
        = \bar{v}_\sigma(p) \qquad \text{(initial)} \\[0.3cm]
         &\vcenter{\hbox{\includegraphics[width=2cm]{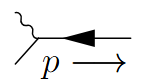}}}
        \quad
        = v_\sigma(p) \qquad \text{(final)}
    \end{cases} \label{eq:external_antifermions} \\[0.2cm]
    &\text{External photons:}\qquad
    &&\begin{cases}
         &\vcenter{\hbox{\includegraphics[width=2cm]{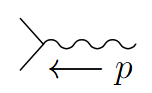}}}
        \quad
        = \epsilon_\lambda(p) \qquad \text{(initial)} \\[0.3cm]
         &\vcenter{\hbox{\includegraphics[width=2cm]{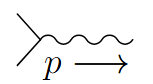}}}
        \quad
        = \epsilon^*_\lambda(p) \qquad \text{(final)}
    \end{cases} \label{eq:external_photons}
\end{alignat}

Here, $Q\in\mathbb{Z}$ is the particle's electric charge in units of the elementary charge unit $e$, such that $Q=\pm1$ for positrons(electrons). Virtual photons and fermions have momentum space amplitudes $\smash{\tilde{D}_{\mu\nu}(q)}$ and $\smash{\tilde{G}}(q)$ [defined in Eqs~(\ref{eq:D_munu}, \ref{eq:G})]. Fermions entering(exiting) the scattering event contribute to the total scattering amplitude with their Dirac spinor $u_\sigma$($\bar{u}_\sigma$) [shown in Eq.~(\ref{eq:external_fermions})], whereas an incoming(outgoing) positron contributes with $\bar{v}_\sigma$($v_\sigma$) [as in Eq.~(\ref{eq:external_antifermions})]. Photons entering(exiting) the scattering event contribute with $\smash{\epsilon^{(*)}_\lambda}$ and $\lambda\in\{\mathrm{L},\mathrm{R}\}$ [see Eq.~(\ref{eq:external_photons})]. The QED vertex, shown in Eq.~(\ref{eq:vertex}), contains two fermionic lines and one photonic line. If a vertex contains a virtual photon and two external fermions, it is customary to define the momentum space conserved $U(1)$ current $\tilde{\jmath}^\mu(p,p') = \bar{u}_{\sigma'}(p')\,\gamma^\mu\,u_\sigma(p)$ with initial and final handednesses $\sigma$ and $\sigma'$. For example, the $t$-channel M\o ller scattering diagram:
\vspace{0.2cm}
\begin{align}\label{eq:t-channel}
    \i\mathcal{M}_t = 
    \vcenter{\hbox{\includegraphics[width=3cm]{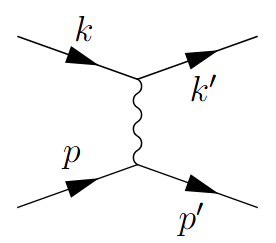}}}
\end{align} \\
The amplitude is proportional to the contraction $\i\mathcal{M}_t \propto \tilde{\imath}_\mu(k,k')\,\tilde{\jmath}^\mu(p,p')$, since the metric tensor in the photon Green's function contracts the electronic $U(1)$ currents $\tilde{\imath}_\mu$ and $\tilde{\jmath}_\mu$, where $\tilde{\imath}_\mu(k,k') \equiv \bar{u}_{\tau'}(k')\,\gamma^\mu\,u_\tau(k)$. Here, $\smash{\tau^{(\prime)}}$ is the initial(final) handedness of the second electron.

Often, the total scattering amplitude consists of multiple channels. For example, M\o ller scattering has a $t$-channel [depicted in Eq.~(\ref{eq:t-channel})], but also a $u$-channel with amplitude $\mathcal{M}_u$, in which the exiting electrons are exchanged. Since electrons are fermions, their fields satisfy anticommutation relations [see Eq.~(\ref{eq:anticommutation})]. Therefore, the total amplitude is given by the difference between the amplitudes of the two channels: $\mathcal{M} = \mathcal{M}_t - \mathcal{M}_u$. Photons, which are bosons, have commuting field operators [see Eq.~(\ref{eq:commutation})]. Therefore when two diagrams differ by an exchange or permutation of photons, the total amplitude is the sum of the individual channels. Examples of processes that involve photon exchange are Compton scattering and electron-positron annihilation into two photons (demonstrated in Section~\ref{subsec:electron-positron}).

\subsection{Lorentz transformations}\label{subsec:Lorentz_transformation}
\texttt{QEDtool} is equipped with functions that Lorentz transform 4-vectors, fields and quantum states. For fields, we differentiate between two Lorentz transformation representations; the $\smash{(1/2,1/2)}$ representation for 4-vectors and the $\smash{(1/2,0)\oplus(0,1/2)}$ representation for Dirac spinors. The generators for the vector representation are~\cite{PeskinSchroeder}
\begin{equation*}
    (\mathcal{J}^{\mu\nu})_{\alpha\beta} = \i\big(\delta^\mu_{\hphantom{\mu}\alpha}\delta^\nu_{\hphantom{\mu}\beta} - \delta^\mu_{\hphantom{\mu}\beta}\delta^\nu_{\hphantom{\mu}\alpha}\big)\,.
\end{equation*}
General 4-tensor Lorentz transformation matrices are then of the form 
\begin{equation}\label{eq:exponentiation}
    \Lambda_{\alpha\beta} = \exp\bigg[ {-\frac{\i}{2}\omega_{\mu\nu}(\mathcal{J}^{\mu\nu})_{\alpha\beta}} \bigg]\,,
\end{equation}
(sometimes abbreviated to $\Lambda$) where $\omega_{\mu\nu}$ is an antisymmetric Lorentz tensor. For pure spatial rotations, $\omega_{0\mu} = 0$ and $\omega_{ij}$ denotes the angles of rotation in the $ij$-plane. For boosts, $\omega_{ij} = 0$ and $\omega_{0j} = \eta_j$ are the rapidity components of the rapidity vector $\boldsymbol{\eta}$. The magnitude $\eta \equiv |\boldsymbol{\eta}|$ is related to the Lorentz factor as $\eta = \arccos\gamma_{\boldsymbol{\beta}}$ where $\gamma_{\boldsymbol{\beta}}$ is the Lorentz factor as defined in Section~\ref{subsec:units}. For the $\smash{(1/2,0)\oplus(0,1/2)}$ representation, the generators are~\cite{PeskinSchroeder}
\begin{equation*}
    S^{\mu\nu} = \frac{\i}{4}\big[ \gamma^\mu, \gamma^\nu \big]\,,
\end{equation*}
and we denote the transformation matrices as $\Lambda_{1/2}$ [replacing $\mathcal{J}^{\mu\nu}$ with $S^{\mu\nu}$ in Eq.~(\ref{eq:exponentiation})]. The gamma matrices connect the two representations through the fundamental property~\cite{PeskinSchroeder,Schwartz,Weinberg}
\begin{equation*}
    \Lambda^{-1}_{1/2}\gamma^\mu\Lambda_{1/2} = \Lambda^\mu_{\hphantom{\mu}\nu}\gamma^\nu\,.
\end{equation*}

\texttt{QEDtool} performs active Lorentz transformations. 4-vectors transform as $p^\mu \to \Lambda^\mu_{\hphantom{\mu}\nu}p^\nu$. The coordinate space Dirac field transforms according to $\psi(x) \to \Lambda_{1/2}\psi(\Lambda^{-1}x)$ and for the Dirac adjoint $\smash{\bar{\psi}(x) \to \bar{\psi}(\Lambda^{-1}x)\Lambda^{-1}_{1/2}}$. For momentum space Dirac spinors, $u(p) \to \Lambda_{1/2}u(p)$ and $\bar{u}(p) \to \bar{u}(p)\Lambda^{-1}_{1/2}$. Because the photon propagator is a rank 2 Lorentzian tensor, it transforms in momentum space as $\smash{\tilde{D}_{\mu\nu}(q) \to \Lambda^\rho_{\hphantom{\mu}\mu}\Lambda^\sigma_{\hphantom{\mu}\nu}\tilde{D}_{\rho\sigma}(q)}$. However, the Lorentz indices in the photon propagator are from the metric tensor, which has components that are invariant under Lorentz transformations. The massive spin-$1/2$ propagators contain 4-vector contractions with gamma matrices in their nominator, whereas their denominators are invariants. Hence they are rank 2 tensors with spinor indices, which transform according to $\tilde{G}(q) \to \Lambda_{1/2}\tilde{G}(q)\Lambda^{-1}_{1/2}$.

In quantum scattering, the QED Feynman amplitudes are Lorentz invariant. However, the quantum states themselves Lorentz transform. We consider two classes of particles: massless gauge bosons (photons) and massive Dirac fermions. Their quantum states have different Lorentz group representations (see Ref.~\cite{Weinberg}), which we will outline here. The standard (rest frame) 4-momentum of the aforesaid fermions is $k = (m,\mathbf{0})$. From here, the more general 4-momentum $p = (\varepsilon_\mathbf{p},\mathbf{p})$ is generated through the Lorentz transformation $p^\mu = L^\mu_{\hphantom{\mu}\nu}(p)\,k^\nu$. Single-particle spin-$s$ states with definite 4-momentum and magnetic projection (along the $z$-axis) $m_s\in\{-s,-s+1,...,s-1,s\}$ can then be defined as $|p,s,m_s\rangle = U[L(p)]\,|k,s,m_s\rangle$. A Lorentz transformation $\Lambda$ on the state $|p,s,m_s\rangle$, is given by the unitary operation~\cite{Weinberg}
\begin{equation}\label{eq:spin-z_boost}
    U(\Lambda)\,|p,s,m_s\rangle = \sum_{m'_s} \mathcal{D}^{(s)}_{m_s'm_s}(W)\,|\Lambda p,s,m'_s\rangle\,,
\end{equation}
where $W \equiv L^{-1}(\Lambda p)\Lambda L(p)$ is an element of the little group, SO(3) for the massive case. $\smash{\mathcal{D}^{(s)}_{m'_s m_s}(W) = \langle p,s,m'_s|\exp(-\i\mathbf{S}\cdot\boldsymbol{\phi})|p,s,m_s\rangle}$ are the Wigner $D$-matrix elements, which form a $(2s+1)$-dimensional irreducible representation of SO(3). Here $\mathbf{S}$ is the spin operator, and the rotation is performed around the axis $\boldsymbol{\phi}/|\boldsymbol{\phi}|$ by an angle $|\boldsymbol{\phi}|$. From here on, we solely consider massive $s=1/2$ particles. These rotate under the 2-dimensional irreducible representation of SU(2). As \texttt{QEDtool} operates in the helicity basis, we define helicity eigenstates by rotating the spin quantization axis ($\hat{\mathbf{z}}$) towards $\hat{\mathbf{p}} = (\cos\phi_p\sin\theta_p, \sin\phi_p\sin\theta_p, \cos\theta_p)$;
\begin{equation}\label{eq:helicity-z_relations}
\begin{aligned}
    &|p,\mathrm{L}\rangle = \exp(\i\phi_p/2)\cos(\theta_p/2)\,|p,\downarrow\rangle - \exp(-\i\phi_p/2)\sin(\theta_p/2)\,|p,\uparrow\rangle\,, \\[0.2cm]
    &|p,\mathrm{R}\rangle = \exp(\i\phi_p/2)\sin(\theta_p/2)\,|p,\downarrow\rangle + \exp(-\i\phi_p/2)\cos(\theta_p/2)\,|p,\uparrow\rangle\,,
\end{aligned}
\end{equation}
which are eigenstates of $\hat{\mathbf{p}}\cdot\mathbf{S}\,|p,\sigma\rangle = (\sigma/2)|p,\sigma\rangle$ with handedness $\sigma$. Lorentz transformations of such helicity eigenstates are of the form $U(\Lambda)\,|p,\sigma\rangle = c_\downarrow U(\Lambda)\,|p,\downarrow\rangle + c_\uparrow U(\Lambda)\,|p,\uparrow\rangle$, which is a superposition of spin-$z$ eigenstates $|\Lambda p, m_s\rangle$. Here $\uparrow\!(\downarrow)$ coincides with the magnetic projections $\smash{m_s = \pm1/2}$. By inverting the relations in Eq.~(\ref{eq:helicity-z_relations}), one can express $U(\Lambda)\,|p,\sigma\rangle$ as a superposition of helicity eigenstates. For example, in Lorentz transforming a left-handed electron, the helicity eigenstates mix as
\begin{equation*}
    U(\Lambda)\,|p,\mathrm{L}\rangle = \Upsilon_\mathrm{L}(\Lambda,p)\,|\Lambda p,\mathrm{L}\rangle + \Upsilon_\mathrm{R}(\Lambda,p)\,|\Lambda p,\mathrm{R}\rangle\,,
\end{equation*}
where we defined the mixing coefficients (derived in \ref{ap:coeffs})
\begin{equation*}
    \begin{aligned}
        \Upsilon_\mathrm{L}(\Lambda,p) &\equiv \cos(\theta_p/2)\,\Big[ \mathcal{D}_{\downarrow\downarrow}(W)\exp[\i(\phi_p-\phi_{\Lambda p})/2] \cos(\theta_{\Lambda p}/2) \\
        &\hspace{2.5cm}- \mathcal{D}_{\uparrow\downarrow}(W)\exp[\i(\phi_p+\phi_{\Lambda p})/2]\sin(\theta_{\Lambda p}/2) \Big] \\
        &\hspace{0.5cm}- \sin(\theta_p/2)\,\Big[ \mathcal{D}_{\downarrow\uparrow}(W)\exp[-\i(\phi_p+\phi_{\Lambda p})/2] \cos(\theta_{\Lambda p}/2) \\
        &\hspace{3cm} - \mathcal{D}_{\uparrow\uparrow}(W)\exp[\i(\phi_{\Lambda p}-\phi_p)/2]\sin(\theta_{\Lambda p}/2)\Big]\,, \\[0.2cm]
        \Upsilon_\mathrm{R}(\Lambda,p) &\equiv \cos(\theta_p/2)\,\Big[\mathcal{D}_{\uparrow\downarrow}(W)\exp[\i(\phi_p+\phi_{\Lambda p})/2] \cos(\theta_{\Lambda p}/2) \\
        &\hspace{2.5cm} + \mathcal{D}_{\downarrow\downarrow}(W)\exp[\i(\phi_p-\phi_{\Lambda p})/2]\sin(\theta_{\Lambda p}/2)\Big] \\
        &\hspace{0.5cm} - \sin(\theta_p/2)\,\Big[ \mathcal{D}_{\uparrow\uparrow}(W)\exp[\i(\phi_{\Lambda p}-\phi_p)/2] \cos(\theta_{\Lambda p}/2) \\
        &\hspace{3cm} + \mathcal{D}_{\downarrow\uparrow}(W)\exp[-\i(\phi_p+\phi_{\Lambda p})/2] \sin(\theta_{\Lambda p}/2)\Big] \,.
    \end{aligned}
\end{equation*}
A similar transformation rule for right-handed electrons can be written. Here, we decomposed $W\in\mathrm{SO}(3)$ into $W = R_z(\alpha)\,R_y(\beta)\,R_z(\gamma)$, which has matrix elements~\cite{Sakurai}
\begin{equation}\label{eq:d-matrix}
    \mathcal{D}^{(1/2)}_{m'_s m_s}(\alpha,\beta,\gamma) = 
    \begin{pmatrix}
        \exp[-\i(\alpha + \gamma)/2]\cos(\beta/2) & -\exp[-\i(\alpha - \gamma)/2]\sin(\beta/2) \\[0.15cm]
        \exp[\i(\alpha - \gamma)/2]\sin(\beta/2) & \exp[\i(\alpha + \gamma)/2]\cos(\beta/2)
    \end{pmatrix}\,.
\end{equation}
$\alpha$, $\beta$ and $\gamma$ are referred to as the Euler angles.

Photons have a standard 4-momentum $k = (\kappa, 0, 0, \kappa)$ with $\kappa\in\mathbb{R}$, whose little group is ISO(2), with elements that can be decomposed into a $z$-rotation and a Lorentz transformation; $W(\delta,\zeta,\theta) = H(\delta,\zeta)\,R(\theta)$~\cite{Weinberg}. Here, $H(\delta,\zeta)$ is a Lorentz transformation that leaves $k$ invariant, parameterized by $\delta$ and $\zeta$, while $R(\theta)$ is a rotation around the $z$-axis by an angle $\theta$. The Lorentz transformation on a photon's momentum-helicity eigenstate $|p,\lambda\rangle$ with helicity $\lambda = \pm1$, is given by~\cite{Weinberg}
\begin{equation*}
    U(\Lambda)\,|p,\lambda\rangle = \exp(\i\lambda\theta)\,|\Lambda p,\lambda\rangle\,.
\end{equation*}
In contrast to the massive spin-$1/2$ representation, the photon helicity eigenstates do not mix, as a consequence of its Lorentz invariance. This property is intrinsically related to the fact that a photon has no rest frame.

\subsection{Entanglement and polarization}\label{subsec:concurrence_Stokes}

The evaluation of entanglement and correlations in the polarization/helicity degrees of freedom is the main deliverable of our package.

Because both photons and massive spin-$1/2$ fermions have internal degrees of freedom with Hilbert spaces of dimension $2$, any such pair forms a two-qubit system. With \texttt{QEDtool}, one can calculate the concurrence of such a two-qubit system. 

The concurrence is an entanglement monotone -- i.e., a quantity which cannot grow under local operations and classical communication -- introduced by Wootters in the late nineties~\cite{Wootters1998}, and is a standard {\em bona fide} measure of entanglement for two qubits.

For a two-qubit pure state $|\psi\rangle$, the spin-flipped conjugate is defined as $|\tilde{\psi}\rangle \equiv (\sigma_2 \otimes \sigma_2)|\psi^*\rangle$ with $\sigma_2$ the $y$-Pauli operator and the superscript asterisk denoting the complex conjugate. When $|\psi\rangle$ is normalized to unity, then the concurrence $C \in [0,1]$ of $|\psi\rangle$ is defined as
\begin{equation}\label{eq:concurrence}
    C(\psi) \equiv \big|\langle\psi|\tilde{\psi}\rangle\big|\,.
\end{equation}
Then, $C = 1$ corresponds to a maximally entangled (``Bell'') state, where the local entropy of each individual qubit is maximal, whereas $C = 0$ implies that the state is a fully separable product state (a ``separable'' state being one that is not entangled). Eq.~(\ref{eq:concurrence}) can be carried over to the density matrix $\rho$ of a mixed state such as Eq.~(\ref{eq:projected-out}), whose spin-flipped conjugate reads
\begin{equation*}
    \tilde{\rho} = (\sigma_2 \otimes \sigma_2)\rho^*(\sigma_2 \otimes \sigma_2)\,,
\end{equation*}
and the concurrence would be
\begin{equation*}
    C(\rho) = \max\!\Big( 0, \sqrt{\lambda_1} - \sqrt{\lambda_2} - \sqrt{\lambda_3} - \sqrt{\lambda_4} \Big)\,,
\end{equation*}
where $\lambda_j$ are the eigenvalues of $Q \equiv \rho\tilde{\rho}$ in descending order.

In classical and quantum optics, the polarization state of light is characterized by Stokes parameters~\cite{chandrasekhar,jackson}, which we shall carry over to the helicity of electrons and positrons. Classically, the Stokes parameters are four real numbers $\smash{S_\mu}$ with $\mu \in \{0, 1, 2, 3\}$. Here, $\smash{S_0}$ is the total intensity, $\smash{S_1}$ signifies the intensity difference between horizontal and vertical polarization, $\smash{S_2}$ is the intensity difference between diagonal and antidiagonal polarization, and $S_3$ is the intensity difference between the two circular polarizations~\cite{jackson}. From the Stokes vector $\mathbf{S} = (S_1, S_2, S_3)$, the degree of polarization is defined as
\begin{equation*}
    P_{(1)} = |\mathbf{S}|
\end{equation*}
is the degree of polarization~\cite{Stokes, MultiphotonStokes}. For $P_{(1)} = 1$, the light source is said to be fully polarized. $P_{(1)} = 0$ implies a totally unpolarized source.

Quantum mechanically, the single-particle Stokes parameters from the helicity basis are the expectation values of the Pauli operators. This notion can be extended to multi-particles states. For $n$-particle states, the Stokes tensor equals (see e.g. Refs.~\cite{MultiphotonStokes,measure_qubits})
\begin{equation*}
    S_{\mu_1 \cdots\, \mu_n} = \mathrm{tr}\,\Bigg[ \rho\, \bigotimes_{j=1}^n \sigma_{\mu_j} \Bigg]\,.
\end{equation*}
The two-particle Stokes parameters may be interpreted as follows: the correlation $S_{11} = -1$ would imply that one particle is horizontally polarized and the other vertically, or {\em vice versa}. Off-diagonal Stokes parameters, say, e.g., $S_{23}$, heuristically represent the degree (correlation) to which one particle is, say, diagonally polarized while the other is circularly polarized. Note that, in quantum mechanics, the two-particle degree of polarization is not well-defined by a quantity such as $(S_{11}^2 + S_{22}^2 + S_{33}^2)^{1/2}$. Namely, there exist two-particle states where the first moments of the Stokes operators are zero, whereas e.g. the variances are not, referred to as hidden polarizations~\cite{HiddenPolarizations1, HiddenPolarizations2}. 

In earlier work~\cite{Stokes}, it was shown that correlations of the form $S_{0j}$ and $S_{j0}$ are single-particle degrees of polarizations, and that they play a role in determining the entanglement of the state. They characterize the degree in which the particles behave as a pair and they are zero for a maximally entangled two-qubit system. By exploiting this behavior, the so-called two-particle degree of polarization can be defined as \cite{Stokes}
\begin{equation*}
    P_{(2)} \equiv 1 - \frac{1}{2}\sum_{j=1}^3\Big(S^2_{0j} + S^2_{j0}\Big)\,.
\end{equation*}
$P_{(2)}$ is related to entanglement in the sense that 
that highly entangled states are accompanied by a high degree of polarization.

\section{Usage and examples}\label{sec:examples}

\texttt{QEDtool} was developed and tested in \texttt{python 3.11.5}, and it makes use of the third-party libraries \texttt{numpy} and \texttt{transforms3d}~\cite{transforms3d}. It is released under the MIT license, and it can be cloned from the GitHub repository~\cite{GitHub}, which also includes Jupyter notebooks that contain examples. Moreover, it can be installed directly from PyPI by running the following command in the terminal:
\begin{Verbatim}
    $: pip install qedtool 
\end{Verbatim}

In this section, we provide various examples, ranging from vector and quantum state operations to complete scattering processes. For all examples in this section, we made the imports
\begin{Verbatim}
    import numpy as np
    import qedtool as qtl
\end{Verbatim}
Sections \ref{subsec:3-vector} and \ref{subsec:4-vector} contain basic examples of 3- and 4-vectors. In Section~\ref{subsec:quantum_states} we define quantum states and we calculate their concurrence, Stokes parameters and degree of polarization. We present few examples in which we demonstrate the effects of Lorentz transformations on quantum states. In Section~\ref{subsec:particles_boost}, we present an example in which we define a \texttt{RealParticle} instance, which we Lorentz transform. We include two scattering examples, one of which makes use of the standard \texttt{QEDtool} modules; electron-positron annihilation into photons (Section~\ref{subsec:electron-positron}). We will initially describe the annihilation from the CM frame, after which we boost to a moving reference frame. The other example, outlined in Section~\ref{subsec:Bhabha_diagrams}, makes use of the \texttt{standard\_scattering} function. We demonstrate the \texttt{standard\_scattering} function applied to Bhabha scattering and how it can be used to evaluate emitted states.

\subsection{3-vectors and basic operations}\label{subsec:3-vector}
Within the context of relativistic scattering, the \texttt{ThreeVector} class is mainly used to define 3-vectors such as boost vectors $\boldsymbol{\beta}$ for boosts and angle vectors $\boldsymbol{\theta}$ for rotations. With the following commands:
\begin{Verbatim}
    >>> u = qtl.ThreeVector(-1, -7, 2, "Cartesian")
    >>> v = qtl.ThreeVector(1, 1.57, 0.1, "spherical")
    >>> w = qtl.ThreeVector(2.7, 0.2, 0.36, "cylindrical")
\end{Verbatim}
we create 3-vectors $\mathbf{u}$, $\mathbf{v}$ and $\mathbf{w}$, by specifying the Cartesian, spherical, and cylindrical components respectively. As an \texttt{ndarray},
\begin{Verbatim}
    >>> print(w.vector)
    [2.64617976 0.53640719 0.36      ]
\end{Verbatim}
By running the command \texttt{w.sphericals} we get the spherical components of \texttt{w}:
\begin{Verbatim}
    >>> print(w.sphericals)
    (2.7238942710758804, 1.4382447944982226, 0.2)
\end{Verbatim}
As mentioned, users can create linear combinations of 3-vectors using the standard Python syntax
\begin{Verbatim}
    >>> x = u - 3.2 * v + 11 * w
    >>> print(x.vector)
    [24.92396504 -1.41898771  5.95745175]
\end{Verbatim}
The Euclidean inner product is simply calculated as
\begin{Verbatim}
    >>> a = qtl.ThreeVector(2, 1, 0.6, "spherical")
    >>> print(a * a)
    4.000000000000001
\end{Verbatim}
\texttt{-a} flips the direction of the 3-vector. 

Both the \texttt{ThreeVector.dot} and \texttt{ThreeVector.beta} methods will return an instance of \texttt{ThreeVector}. As an example, we will multiply an \texttt{ndarray} of shape \texttt{(3, 3)} with the previously-defined 3-vector \texttt{a}:
\begin{Verbatim}
    >>> matrix = np.array([[3.1, 2.7, 0], 
                           [4.1, 0.5, -1.2], 
                           [-7.7, 0.3, 1.8]])
    >>> b = qtl.ThreeVector.dot(matrix, a)
    >>> print(b.vector)
    [ 6.87157842,  4.8732717 , -8.46507152]
\end{Verbatim} 
Operations such as \texttt{0.2/u} raise an error.

\subsection{Time-like and light-like 4-vectors}\label{subsec:4-vector}
Constructing and linearly combining 4-vectors is done similarly to 3-vectors (Section \ref{subsec:3-vector}). As an example we will construct an on-shell 4-momentum and boost it with its own sign-flipped boost vector:
\begin{Verbatim}
    >>> m = qtl.constant("electron mass")
    >>> qmu = qtl.FourVector(np.sqrt(1 + m**2), 1, 0, 0)
    >>> beta = qtl.ThreeVector.beta(qmu)
    >>> qmu_b = qtl.boost(qmu, -beta)
    >>> print(qmu_b.vector)
    [ 5.11000000e-01  0.00000000e+00  0.00000000e+00 -1.61666236e-11]
\end{Verbatim}
In other words, the 4-momentum of an electron observed from a co-moving frame is $(m, \mathbf{0})$. For a light-like 4-momentum $k = (|\mathbf{k}|, \mathbf{k})$, it should hold that $|\boldsymbol{\beta}| = 1$,
\begin{Verbatim}
    >>> kmu = qtl.FourVector(7, 7, 0.1, 0.2, "spherical")
    >>> beta_photon = qtl.ThreeVector.beta(kmu)
    >>> print(beta_photon.sphericals)
    (0.9999999999999999, 0.10000000000000056, 0.19999999999999946)
\end{Verbatim}
Notice that $\boldsymbol{\beta} \parallel \mathbf{k}$. We will now boost $k$ and check whether it remains light-like, as it should:
\begin{Verbatim}
    >>> boost_vec = qtl.ThreeVector(0.7, 0.4, 0.1, "spherical")
    >>> kmu_b = qtl.boost(kmu, boost_vec)
    >>> print(kmu_b * kmu_b)
    5.684341886080802e-14
\end{Verbatim}
which is light-like (taking the floating point errors into account). Note here, that using an asterisk to multiply 4-vectors automatically takes the Lorentzian inner product. Moreover, for $\texttt{amu} = (a^0, \mathbf{a})$, \texttt{-amu} returns $(a^0, -\mathbf{a})$.

\subsection{Quantum states, entanglement and polarization}\label{subsec:quantum_states}
We start by creating single-particle polarization states with the same 4-momentum (therefore, we will omit the 4-momenta in the ket-notation until we Lorentz transform quantum states). We define an electronic 4-momentum in the $z$-direction:
\begin{Verbatim}
    >>> m = qtl.constant("electron mass")
    >>> pmu = qtl.FourVector(np.sqrt(1 + m**2), 1, 0, 0)
\end{Verbatim}
Then, we can define single-particle states:
\begin{Verbatim}
    >>> l = qtl.QuantumState.single(pmu, "L")
    >>> r = qtl.QuantumState.single(pmu, "R")
    >>> h = qtl.QuantumState.single(pmu, "H")
    >>> v = qtl.QuantumState.single(pmu, "V")
\end{Verbatim}
The L- and R-polarization states should be orthogonal, i.e. $\langle\mathrm{L}|\mathrm{R}\rangle = 0$:
\begin{Verbatim}
    >>> print(qtl.inner_product(l, r))
    0
\end{Verbatim}
The L- and V-polarization states must have some nonzero overlap; $\langle\mathrm{L}|\mathrm{V}\rangle = \i/\sqrt{2}$ [see Eq.~(\ref{eq:polarization_conventions})]
\begin{Verbatim}
    >>> print(qtl.inner_product(l, v))
    0.7071067811865475j
\end{Verbatim}
Two-particle states can be constructed by taking the tensor product of single particle states. Consider the Bell state $\smash{|\psi\rangle = (|\mathrm{LR}\rangle + |\mathrm{RL}\rangle)/\sqrt{2}}$, which is generated as
\begin{Verbatim}
    >>> psi = (l * r + r * l) / np.sqrt(2)
\end{Verbatim}
Since this is a maximally entangled two-particle state, its concurrence can be calculated and it should equal unity;
\begin{Verbatim}
    >>> print(qtl.concurrence(psi))
    0.9999999999999999
\end{Verbatim}
Additionally, the $S_{33}$ parameter of $|\psi\rangle$ should equal to $-1$ as the particles are entangled in opposite polarization modes;
\begin{Verbatim}
    >>> print(qtl.stokes_parameter(psi, [3, 3]))
    -1.0
\end{Verbatim}
In the $\{\mathrm{H},\mathrm{V}\}$ basis, $\smash{|\psi\rangle = (|\mathrm{HH}\rangle + |\mathrm{VV}\rangle)/\sqrt{2}}$. This reveals that $S_{11} = 1$, since the superposition consists of equal polarization modes in the $\{\mathrm{H},\mathrm{V}\}$ basis:
\begin{Verbatim}
    >>> print(qtl.stokes_parameter(psi, [1, 1]))
    1.0
\end{Verbatim}
The state $|\psi\rangle$ is of course highly polarized, hence the degree of two-particle polarization should be unity as well:
\begin{Verbatim}
    >>> print(qtl.degree_polarization(psi))
    1.0
\end{Verbatim}
The \smash{\texttt{stokes\_parameter}} function can also be used to calculate Stokes parameters for $n$-particle states with $n > 2$. Here we provide a $4$-particle Stokes parameter example, with the quantum state $|\Phi\rangle = ( |\mathrm{LRHV}\rangle - |\mathrm{LLRR}\rangle + |\mathrm{RHHL}\rangle - |\mathrm{HVHV}\rangle )/2$. This is calculated with the commands
\begin{Verbatim}
    >>> phi_4pcl = (l * r * h * v - l * l * r * r \
                    + r * h * h * l - h * v * h * v) / 2
    >>> print(qtl.stokes_parameter(phi_4pcl, [3, 1, 2, 1]))
    0.2222222222222223
\end{Verbatim}

We will now study the effect of a boost on a two-particle state. Consider the two-particle electronic state $|\Psi\rangle = |p_+,\mathrm{L};p_-,\mathrm{R}\rangle$ with $p_\pm \equiv (\varepsilon_\mathbf{p}, \pm\mathbf{p})$. For this, we define
\begin{Verbatim}
    >>> plus_pmu_left = qtl.QuantumState.single(pmu, "L")
    >>> minus_pmu_right = qtl.QuantumState.single(-pmu, "R")
    >>> state_electron_pair = plus_pmu_left * minus_pmu_right
\end{Verbatim}
To boost \texttt{state\_electron\_pair}, we boost the individual states $U(\Lambda)\,|p_\pm,\mathrm{L(R)}\rangle$ with boost vector $\boldsymbol{\beta}$:
\begin{Verbatim}
    >>> beta = qtl.ThreeVector(0.6, 1.4, 0.8)
    >>> plus_pmu_left_b = qtl.boost(plus_pmu_left, beta)
    >>> minus_pmu_right_b = qtl.boost(minus_pmu_right, beta)
\end{Verbatim}
The boosted electron pair state is then constructed as
\begin{Verbatim}
    >>> state_electron_pair_b = plus_pmu_left_b * minus_pmu_right_b
\end{Verbatim}
By constructing a $3\times3$ Stokes matrix with elements
\begin{equation*}
    S_{ij} = \texttt{qtl.stokes\_parameters(state\_electron\_pair, [i, j])}
\end{equation*}
(and similarly for the boosted pair) we observe that helicity eigenstates of Dirac fermions mix in a Lorentz boost. The Stokes matrix of \texttt{state\_electron\_pair} reads
\begin{Verbatim}
    [[ 0.  0.  0.]
     [ 0.  0.  0.]
     [ 0.  0. -1.]]
\end{Verbatim}
while that of the boosted electron pair state, \texttt{state\_electron\_pair\_b}, equals
\begin{Verbatim}
    [[ 0.39156398  0.         -0.6215289 ]
     [ 0.          0.          0.        ]
     [ 0.71328604  0.         -0.604642  ]]
\end{Verbatim}

\subsection{Real particles}\label{subsec:particles_boost}
In this section, we provide an example in which we work with particles and boosts thereon. We will create a 200 keV electron, working in keV units. The electron is moving in the negative $z$-direction, hence we define the 4-momentum as
\begin{Verbatim}
    >>> p = 200
    >>> m = qtl.constant("electron mass", "keV")
    >>> pmu = qtl.FourVector(np.sqrt(p**2 + m**2), p, np.pi, 0)
\end{Verbatim}
We let the electron be right-handed, moving towards an interaction vertex;
\begin{Verbatim}
    >>> electron = qtl.RealParticle.electron(1, pmu, "in")
\end{Verbatim}
Printing its properties yields
\begin{Verbatim}
    >>> print(electron.mass)
    511.0
    >>> print(electron.charge)
    -1.0
    >>> print(electron.four_momentum.vector)
    [ 5.48744932e+02  2.44929360e-14  0.00000000e+00 -2.00000000e+02]
    >>> print(electron.polarization.bispinor)
    [1.14349642e-15+0.j 1.86747137e+01+0.j 1.67551301e-15+0.j
     2.73632040e+01+0.j]
\end{Verbatim}
The electron's 4-momentum and bispinor as printed above are as expected; $p = (\varepsilon_\mathbf{p}, 0, 0, -|\mathbf{p}|)$ and $\smash{u_{\mathrm{R}}(p) = \big(0, \sqrt{\varepsilon_\mathbf{p}-|\mathbf{p}|}, 0, \sqrt{\varepsilon_\mathbf{p}+|\mathbf{p}|}\big)}$. Relative to its CM frame, the electron has a velocity
\begin{Verbatim}
    >>> beta_cm = qtl.ThreeVector.beta(electron.four_momentum)
    >>> velocity = np.sqrt(beta_cm * beta_cm)
    >>> print(velocity)
    0.3644680587798214
\end{Verbatim}
Instead of separately boosting its polarization and 4-momentum, one can boost the electron as a whole. Here, we will boost the electron with $-\boldsymbol{\beta}_{\text{CM}}$, retrieving its CM polarization and 4-momentum:
\begin{Verbatim}
    >>> electron_cm = qtl.boost(electron, -beta_cm)
\end{Verbatim}
Its CM 4-momentum and polarization are
\begin{Verbatim}
    >>> print(electron_cm.four_momentum.vector)
    [ 5.11000000e+02  1.29878059e-26  0.00000000e+00 -1.06041398e-10]
    >>> print(electron_cm.polarization.bispinor)
    [1.38417597e-15+0.j 2.26053091e+01+0.j 1.38417597e-15+0.j
     2.26053091e+01+0.j]
\end{Verbatim}
which are approximately $p_{\text{CM}} = (m, \mathbf{0})$ and $u_{\mathrm{R}}(p_{\text{CM}}) = \sqrt{m}\,(0,1,0,1)$.

\subsection{Electron-positron annihilation}\label{subsec:electron-positron}
The example presented in this section is a complete scattering processes, in contrast to the previous sections. We will demonstrate various calculations regarding the electron-positron annihilation into photons: $\e^+\e^- \to 2\gamma$. The calculation is performed in the CM frame, where the electron and positron have 4-momenta $p_\pm = (\varepsilon_\mathbf{p}, 0, 0, \pm|\mathbf{p}|)$. Due to 4-momentum conservation, the emitted photons must have 4-momenta $k_{1}$ and $k_2$ that are in opposite directions with $k^0_{1,2} = |\mathbf{k}_{1,2}| = \varepsilon_\mathbf{p}$. Within this setting, we want to calculate the differential cross section, concurrence and Stokes parameters for some specified initial state. For the initial state, we define the mixed state which is a convex combination [see Eq.~(\ref{eq:convex})] of the pure states
\begin{equation}\label{eq:mixed_example}
\begin{aligned}
    |\psi_1\rangle &= \frac{1}{\sqrt{2}}\Big( |p_+,\mathrm{L};p_-,\mathrm{R}\rangle + |p_+,\mathrm{H};p_-,\mathrm{V}\rangle \Big)\,, \\[0.2cm]
    |\psi_2\rangle &= \vphantom{\frac{1}{1}}|p_+,\mathrm{R};p_-,\mathrm{R}\rangle\,,
\end{aligned}
\end{equation}
with $w_1 = 2/5$ and $w_2 = 3/5$.
\begin{enumerate}
    \item We start off by setting the energy units. We choose MeV, as the electron mass is on the MeV scale. Additionally, we denote the electron mass and charge by \texttt{m} and \texttt{e} respectively. For this we run the commands
    \begin{Verbatim}
    m = qtl.constant("electron mass")
    e = qtl.constant("elementary charge")
    \end{Verbatim}
    Moreover, we will fix the magnitude of the initial electron 3-momentum, e.g. $|\mathbf{p}| = 0.5~\mathrm{MeV}$, while scanning over the full polar angle, fixing $\phi$:
    \begin{Verbatim}
    p = 0.5
    phi = 0
    theta = np.linspace(0, np.pi, 200)
    energy = np.sqrt(p**2 + m**2)
    \end{Verbatim}

    \item We can already define the initial electron and positron 4-momenta,
    \begin{Verbatim}
    pmu_e = qtl.FourVector(energy, p, 0, 0)
    pmu_p = -pmu_e
    \end{Verbatim}
    and their initial quantum state. Firstly, the single-particle momentum eigenstates:
    \begin{Verbatim}
    plus_l = qtl.QuantumState.single(pmu_e, "L")
    plus_r = qtl.QuantumState.single(pmu_e, "R")
    plus_h = qtl.QuantumState.single(pmu_e, "H")
    minus_r = qtl.QuantumState.single(pmu_p, "R")
    minus_v = qtl.QuantumState.single(pmu_p, "V")
    \end{Verbatim}
    The complete mixed initial state [see Eq.~(\ref{eq:mixed_example})], is defined using the commands
    \begin{Verbatim}
    state_1 = (plus_l * minus_r + plus_h * minus_v) / np.sqrt(2)
    state_2 = plus_r * minus_r

    w = [0.4, 0.6]
    states = [state_1, state_2]
        
    in_state = qtl.QuantumState.mixed(states, w)
    \end{Verbatim}
    where the asterisk takes the tensor product. Note that if we considered this computation over a range of $|\textbf{p}|$ values, this step would take place insider a $\texttt{for}$ loop over all $|\textbf{p}|$ values (i.e. would take place in step 4).
    
    \item For computing all 2-to-2 Feynman amplitudes, we will need all handedness configurations, which we will sort according to the photon helicities:
    \begin{Verbatim}
    hand_ll = qtl.handedness_config(4, [2, 3], [-1, -1])
    hand_lr = qtl.handedness_config(4, [2, 3], [-1, 1])
    hand_rl = qtl.handedness_config(4, [2, 3], [1, -1])
    hand_rr = qtl.handedness_config(4, [2, 3], [1, 1])
        
    h = [hand_ll, hand_lr, hand_rl, hand_rr]
    \end{Verbatim}
    Here, \texttt{hand\_ll} contains all handedness configurations corresponding to the emitted photon state $|\mathrm{LL}\rangle$. In total, we will calculate twelve quantities; the differential cross section, the concurrence, the degree of two-photon polarization, and nine Stokes parameters. Therefore we need twelve empty lists. These empty lists are all contained within an overarching list named \texttt{data}:
    \begin{Verbatim}
    data = qtl.empty_lists(12)
    \end{Verbatim}
    
    \item We loop over the polar angle and over all handedness configurations. For every angle, we construct the amplitude matrix, i.e. the $4\times4$ matrix that stores the Feynman amplitudes. We denote this matrix by \texttt{amplitudes}.
    \begin{Verbatim}
    for i in range(len(theta)):
        amplitudes = []
        for j in range(len(h)):
            amplitudes_row = []
            for k in range(len(h[j])):
                # k-loop
                ...
    \end{Verbatim}
    Here \texttt{amplitudes\_row} is the row of \texttt{amplitudes}, which will be filled in the innermost handedness loop. The \texttt{amplitudes} matrix is of the form
    \begin{equation}\label{eq:amplitudes}
        \texttt{amplitudes} = \begin{pmatrix}
            \mathcal{M}_{\mathrm{LL}\to\mathrm{LL}} & \mathcal{M}_{\mathrm{LL}\to\mathrm{LR}} & \mathcal{M}_{\mathrm{LL}\to\mathrm{RL}} & \mathcal{M}_{\mathrm{LL}\to\mathrm{RR}} \\
            \mathcal{M}_{\mathrm{LR}\to\mathrm{LL}} & \mathcal{M}_{\mathrm{LR}\to\mathrm{LR}} & \mathcal{M}_{\mathrm{LR}\to\mathrm{RL}} & \mathcal{M}_{\mathrm{LR}\to\mathrm{RR}} \\
            \mathcal{M}_{\mathrm{RL}\to\mathrm{LL}} & \mathcal{M}_{\mathrm{RL}\to\mathrm{LR}} & \mathcal{M}_{\mathrm{RL}\to\mathrm{RL}} & \mathcal{M}_{\mathrm{RL}\to\mathrm{RR}} \\
            \mathcal{M}_{\mathrm{RR}\to\mathrm{LL}} & \mathcal{M}_{\mathrm{RR}\to\mathrm{LR}} & \mathcal{M}_{\mathrm{RR}\to\mathrm{RL}} & \mathcal{M}_{\mathrm{RR}\to\mathrm{RR}}
        \end{pmatrix}\,.
    \end{equation}
    
    \item Now that we defined the angular grid and the necessary arrays to save the data, we formulate the kinematics and solve for the Feynman amplitudes. We will stay within the \texttt{k}-loop until step 7. Starting with the 4-momenta:
    \begin{Verbatim}
    # k-loop
    kmu_1 = qtl.FourVector(energy, energy, theta[i], phi)
    kmu_2 = -kmu_1
    \end{Verbatim}
    As the scattering process is described in the CM frame, we only define \texttt{pmu} for the electron and \texttt{kmu} for the first photon. The 4-momenta of the positron and second photon are then simply \texttt{-pmu} and \texttt{-kmu}. Since the Euclidean norms of the photon 3-momenta are constant, it is customary to define their 4-momenta using spherical coordinates. Here we did not specify \texttt{coordinates} as it is set to \texttt{"spherical"} by default.
    
    \item The two first-order tree-level pair annihilation diagrams are
    \vspace{0.2cm}
    \begin{align*}
        \vcenter{\hbox{\includegraphics[width=3.3cm]{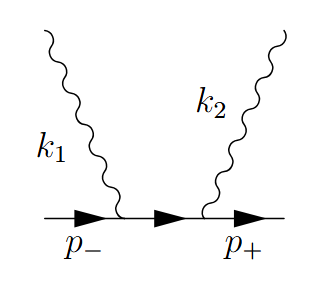}}} \,\,\quad+\,\,\quad
        \vcenter{\hbox{\includegraphics[width=3.3cm]{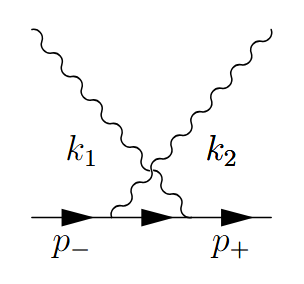}}} \\[-0.15cm]
    \end{align*}
    To define two internal fermionic states, we write the lines
    \begin{Verbatim}
    qmu_1 = pmu_e - kmu_1
    qmu_2 = pmu_e - kmu_2
    \end{Verbatim}
    
    \item After specifying all momenta and real particle helicities, we are in the position to construct the particles themselves. Firstly, the real particles:
    \begin{Verbatim}
    electron = qtl.RealParticle.electron(h[j][k][0], pmu_e, "in")
    positron = qtl.RealParticle.positron(h[j][k][1], pmu_p, "in")
    photon_1 = qtl.RealParticle.photon(h[j][k][2], kmu_1, "out")
    photon_2 = qtl.RealParticle.photon(h[j][k][3], kmu_2, "out")
    \end{Verbatim}
    Note here that the indices for the photon helicities are 2 and 3, i.e. the fixed helicities in step 3. The electron and positron contribute to the Feynman amplitude with their polarizations $u_\sigma(p_-)$ and $\bar{v}_{\sigma'}(p_+)$. These are attributes of the \texttt{RealParticle} class,
    \begin{Verbatim}
    u = electron.polarization.bispinor
    v = positron.polarization.bispinor
    \end{Verbatim}
    Here \texttt{electron.polarization} is an instance of the class \texttt{DiracSpinor}, therefore \texttt{electron.polarization.bispinor} is an \texttt{ndarray}. Moreover, the Dirac adjoint of the positron's polarization is taken during the construction of the \texttt{RealParticle} instance, as \texttt{direction} was specified to be \texttt{"in"}. Since both photons couple to a vertex, it is conventional to define their polarization matrices $-\i e \slashed{\epsilon}^*_\lambda(k_{1,2})$. This can be done with the following two lines:
    \begin{Verbatim}
    e1 = -1j * e * qtl.slashed(photon_1.polarization)
    e2 = -1j * e * qtl.slashed(photon_2.polarization)
    \end{Verbatim}
    In a similar fashion, we construct the virtual fermionic states as
    \begin{Verbatim}
    fermion_1 = qtl.VirtualParticle.electron(qmu_1)
    fermion_2 = qtl.VirtualParticle.electron(qmu_2)
    \end{Verbatim}
    which contribute to the Feynman amplitude with their propagators;
    \begin{Verbatim}
    g1 = fermion_1.propagator
    g2 = fermion_2.propagator
    \end{Verbatim}
    
    \item The total amplitude, i.e. the sum of the two diagrams, is given by
    \begin{equation*}
        -e^2\bar{v}_{\sigma'}(p_+)\slashed{\epsilon}^*_{\lambda'}(k_2)\tilde{G}(q_1)\slashed{\epsilon}^*_\lambda(k_1)u_\sigma(p_-) -e^2\bar{v}_{\sigma'}(p_+)\slashed{\epsilon}^*_{\lambda}(k_1)\tilde{G}(q_2)\slashed{\epsilon}^*_{\lambda'}(k_2)u_\sigma(p_-)\,,
    \end{equation*}
    which translates to the simple line
    \begin{Verbatim}
    amplitude = v.dot(e2).dot(g1).dot(e1).dot(u) \
                + v.dot(e1).dot(g2).dot(e2).dot(u)
    \end{Verbatim}
    
    After filling the \texttt{amplitudes} matrix,
    \begin{Verbatim}
        ...
        amplitudes_row.append(amplitude)
    amplitudes.append(amplitudes_row)
    \end{Verbatim}
    
    we exit the \texttt{k} and \texttt{j} handedness configuration loops and we are back in the $\theta$-loop.
    \item Here (in the $\theta$-loop) we calculate all of the sought quantities, all of which require the post-scattering quantum state
    \begin{Verbatim}
        ...
    out = qtl.QuantumState.out_state(in_state, amplitudes)
    \end{Verbatim}
    
    i.e. Eq.~(\ref{eq:projected-out}). The differential cross section, concurrence, two-photon degree of polarization and the two-photon Stokes parameters are then obtained through
    \begin{Verbatim}
    dcs = qtl.diff_cross_section(pmu_e, pmu_p, out)
    conc = qtl.concurrence(out)
    pol = qtl.degree_polarization(out)
    s11 = qtl.stokes_parameter(out, [1, 1])
    ...
    s33 = qtl.stokes_parameter(out, [3, 3])
    \end{Verbatim}
    
    \item For each $\theta$, all quantities are saved to the \texttt{data} array;
    \begin{Verbatim}
    quantities = [dcs, conc, pol, s11, ..., s33]
    for l in range(len(data)):
        data[l].append(quantities[l])
    \end{Verbatim}

    \item Finally, \texttt{data} can be saved as a \texttt{pickle} file, by running the following command outside of all \texttt{for}-loops:
    \begin{Verbatim}
    save_data(filename = "annihilation_photons", 
              keys = ["dcs", "c", 
                      "s11", ..., "s33"], 
              data = data)
    \end{Verbatim}
    
\end{enumerate}
 
\begin{figure}[ht!]
    \centering
    \includegraphics{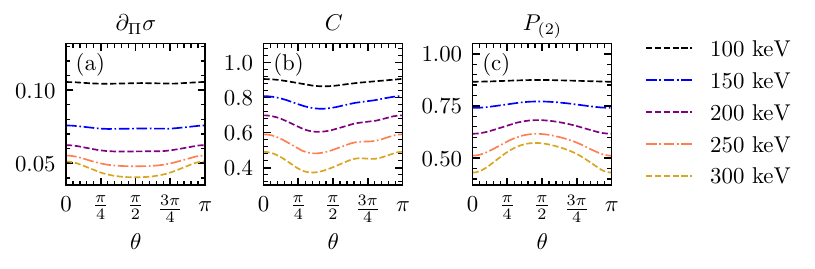}
    \caption{The differential cross section (a), the concurrence (b) and the degree of two-photon polarization (c) of the emitted photon pair created in the annihilation process $\e^+\e^- \to 2\gamma$. The results are computed in the CM frame over an interval of polar angles $\theta\in[0,\pi]$ with $\phi = 0$. Each colored line represents the aforementioned quantities for different collision momenta $|\mathbf{p}|$, ranging from 100~keV to 300~keV.}
    \label{fig:annihilation_prop-conc-DoP}
\end{figure}

Fig.~\ref{fig:annihilation_prop-conc-DoP} contains plots of $\partial_{\Pi}\sigma$, $C$ and $P_{(2)}$ (\texttt{dp}, \texttt{conc} and \texttt{pol}) for various values of $|\mathbf{p}|$, as obtained from the example above. The differential cross section overall decreases with an increasing collision momentum $|\mathbf{p}|$, while the concurrence also decreases. At higher energies, the photon pair is more likely to backscatter or forward scatter. Moreover, a large concurrence is associated with a large two-photon degree of polarization, an one should expect.

The aforementioned quantities can as well be evaluated for smaller numerical steps in $|\mathbf{p}|$. Optionally, by inserting \texttt{progress(idx, array)}, where \texttt{idx} is the index of the outer most loop and array is the looped-over array, the percentage of the completed calculation is printed. 

\begin{figure}[ht!]
    \centering
    \includegraphics{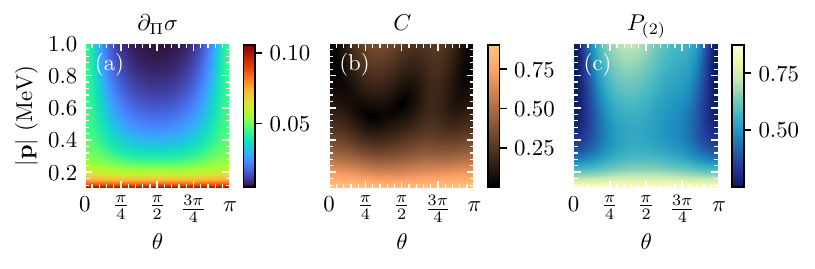}
    \caption{The differential cross section (a), concurrence (b) and two-photon degree of polarization (c) of the photon pair created in the annihilation process $\e^+\e^- \to 2\gamma$. These results are for $\phi = 0$ and an interval of polar angles $\theta\in[0,2\pi]$ and initial electron-positron CM momenta $|\mathbf{p}|\in[0.1,1]$~MeV.}
    \label{fig:annihilation_prop-conc-DoP_continuous}
\end{figure}

Fig.~\ref{fig:annihilation_prop-conc-DoP_continuous} displays the previously computed quantities for a smoother interval of $|\mathbf{p}|$. The differential cross section decreases for all angles with increasing $|\mathbf{p}|$. Around $|\mathbf{p}| \approx 400~\text{keV}$, $\partial_\Pi\sigma|_{\theta=0,\pi}$ seems to decrease less rapidly than $\partial_\Pi\sigma|_{\theta\approx\pi/2}$. At low collision energies, $C \approx 0.9$, and around $|\mathbf{p}| \approx 400$~keV, the concurrence reaches values close to zero. For the higher energies, where $\partial_{\Pi}\sigma|_{\theta\approx\pi/2} \approx 0$, the concurrence and degree of polarization restore to moderate values.
\begin{figure}[ht!]
    \centering
    \includegraphics{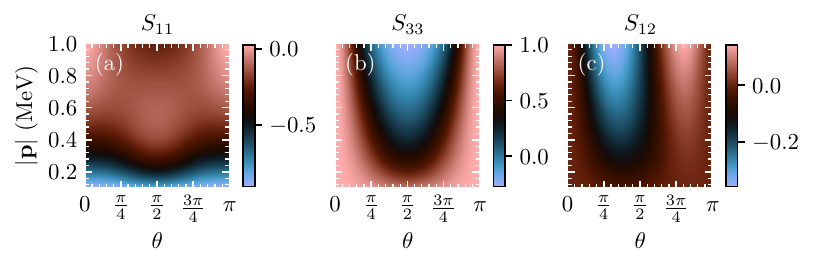}
    \caption{The $S_{11}$, $S_{33}$ and $S_{21}$ two-photon Stokes parameters [(a), (b) and (c) respectively] of the emitted photon pair formed in the annihilation process $\e^+\e^- \to 2\gamma$. The results are for $\phi = 0$, polar angles $\theta\in[0,2\pi]$ and for electron(positron) CM momenta $|\mathbf{p}|\in[0.1,1]$~MeV.}
    \label{fig:annihilation_Stokes}
\end{figure}

In Fig.~\ref{fig:annihilation_Stokes}, the emitted photon pair's two-photon Stokes parameters $S_{11}$, $S_{33}$ and $S_{21}$ are plotted. The degree of double linear (antiparallel) polarization decreases with an increasing collision energy. Interestingly, for all selected collision momenta, the forward scattered and backscattered photon pair is circularly polarized. The $S_{11}$ and $S_{33}$ parameters are symmetric around $\theta = \pi/2$, which is not the case for $S_{21}$. 

\begin{figure}[ht!]
    \centering
    \includegraphics{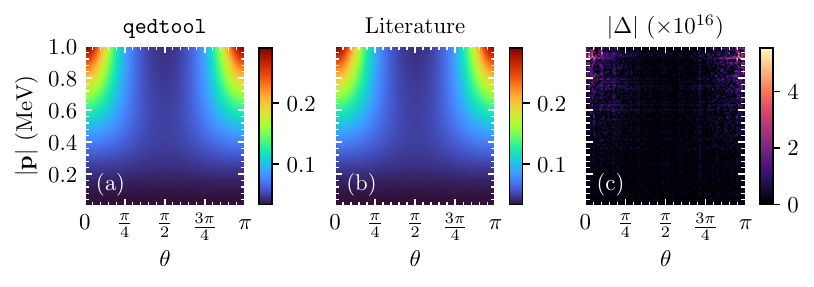}
    \caption{A comparison between the \texttt{QEDtool} and the literary [see Eq.~(\ref{eq:PS_dW})] results [(a) and (b) respectively] of the unpolarized differential scattering probability $\partial_\Pi\mathcal{P}$ of the electron-positron annihilation $\e^+\e^- \to 2\gamma$. Here, $|\Delta|$ denotes the absolute difference between the two results [plotted in (c)], which is on the order of $10^{-16}$.}
    \label{fig:annihilation_literature}
\end{figure}

Here we compute the differential probability $\partial_\Pi\mathcal{P}$ [see Eq.~(\ref{eq:dW/dPi})] of the two-photon state for the unpolarized initial state $\rho = \sum_{\lambda_1,\lambda_2}|\lambda_1\lambda_2\rangle\langle\lambda_1\lambda_2|/4$ with $\lambda_1,\lambda_2\in\{\mathrm{L},\mathrm{R}\}$. Presented as a matrix with polarization indices, it equals the $4\times4$ identity matrix $\smash{\rho_{\alpha\beta} = \delta_{\alpha\beta}/4}$ with $\alpha,\beta\in\{\mathrm{LL},\mathrm{LR},\mathrm{RL},\mathrm{RR}\}$. The analytical expression of the differential scattering probability can be found in standard QFT literature, e.g. Ref.~\cite{PeskinSchroeder}
\begin{equation}\label{eq:PS_dW}
    \frac{\partial\mathcal{P}}{\partial\Pi} = 2e^4\Bigg[ \frac{p_-\cdot k_2}{p_-\cdot k_1} + 2m^2\bigg( \frac{1}{p_-\cdot k_1} + \frac{1}{p_-\cdot k_2} \bigg) - m^4\bigg( \frac{1}{p_-\cdot k_1} + \frac{1}{p_-\cdot k_2} \bigg)^2 \Bigg]\,.
\end{equation}
Fig.~\ref{fig:annihilation_literature} compares the numerical results obtained using \texttt{QEDtool} with Eq.~(\ref{eq:PS_dW}). The absolute difference is on the order of $10^{-16}$, which originates from floating point errors.

\begin{figure}[ht!]
    \centering
    \includegraphics{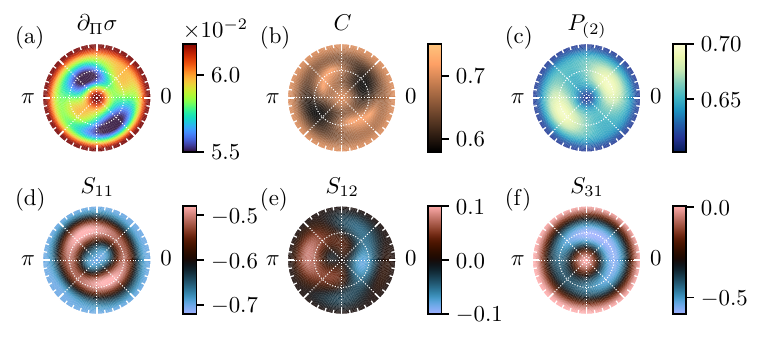}
    \caption{The differential cross section (a), the concurrence (b), the two-photon degree of polarization (c), and three Stokes parameters (d-f) for an interval of polar angles $\theta\in[0,\pi]$, azimuthal angles $\phi \in [0, 2\pi]$, and an initial electron-positron CM momenta $|\mathbf{p}| = 200$~keV. The white dotted circle denotes $\smash{\theta = \pi/2}$.}
    \label{fig:annihilation_all_angular}
\end{figure}

Naturally, one may fix the CM energy, e.g. by fixing $|\mathbf{p}| = 200$~keV, and investigate the solid angle correlation distributions. The calculation of the following results are explained in detail in one of the Jupyter notebooks on our GitHub repository~\cite{GitHub}. Fig.~\ref{fig:annihilation_all_angular} contains differential cross section, concurrence, two-photon degree of polarization, and some Stokes parameters at $|\mathbf{p}| = 200$~keV against the solid angle $(\theta,\phi)$. 

\begin{figure}[ht!]
    \centering
    \includegraphics{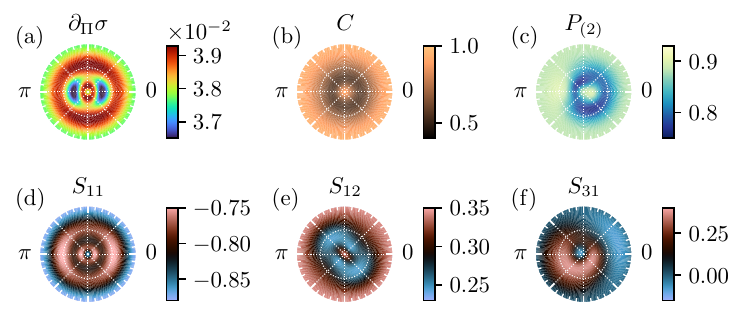}
    \caption{The same quantities as in Fig.~\ref{fig:annihilation_all_angular} from a boosted frame with $\boldsymbol{\beta} = (0, 0, 0.6)$. The interval of the polar angle is $\theta_{\boldsymbol{\beta}}\in[0, \pi]$, the azimuthal angle $\phi_{\boldsymbol{\beta}} \in [0, 2\pi]$, and the electron-positron CM momentum's magnitude is $|\mathbf{p}| = 200$~keV.}
    \label{fig:annihilation_all_angular_boosted}
\end{figure}
We are now in the position to observe the effect of a boost. Feynman amplitudes are Lorentz invariant. However, when boosting to a different reference frame, the definitions of $\theta$ and $\phi$ might change. Therefore, we will boost the outgoing 4-momenta and retrieve the boosted angles $\theta_{\boldsymbol{\beta}}$ and $\phi_{\boldsymbol{\beta}}$. Our boost vector will be in the $z$-direction; $\boldsymbol{\beta} = (0, 0, 0.6)$. We choose the collision axis as the boost direction, otherwise generating solid angle plots (such as Fig.~\ref{fig:annihilation_all_angular}) or plots containing angles and energies (as in Fig.~\ref{fig:annihilation_prop-conc-DoP_continuous}) becomes nontrivial. Namely, $\theta_{\boldsymbol{\beta}}$ will generally depend on $|\Lambda\mathbf{p}|$ and on $\phi$. By picking the collision axis as the boost direction, $\phi_{\boldsymbol{\beta}}$ is independent of $\phi$. Moreover, at $|\mathbf{p}| = 200~\text{keV}$, a boost with $|\boldsymbol{\beta}| = 0.6$ parallel to $\mathbf{p}$ is fast enough to flip the helicity of one of the initial fermions. As a consequence, the two-photon correlations, plotted against the boosted solid angle, is not only ``stretched'' by the transformation of angles; the correlations have transformed due to the Lorentz covariance of the helicity basis. Fig.~\ref{fig:annihilation_all_angular_boosted} shows the boosted two-photon quantities from Fig.~\ref{fig:annihilation_all_angular}.

\subsection{Bhabha scattering and entanglement with \texttt{standard\_scattering}}\label{subsec:Bhabha_diagrams}

\begin{figure}[ht!]
    \centering
    \includegraphics{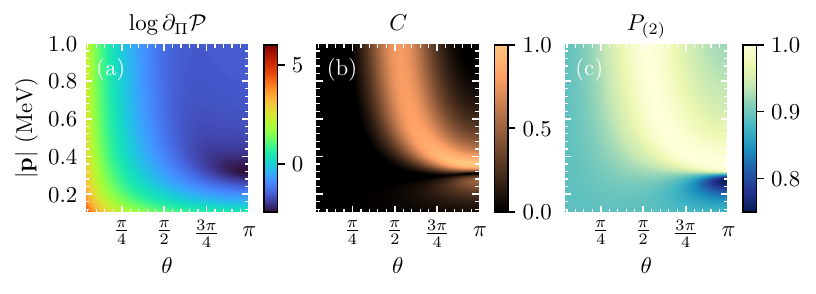}
    \caption{(Logarithm of) the differential scattering probability (a), concurrence (b) and electron-positron pair's degree of polarization (c) for the Bhabha scattered electron-positron pair. The in-state is mixed state, formed by convexly combining the pure states in Eq.~(\ref{eq:bhabha_diag_in_1_2}). The results are computed in the CM frame for $\phi = 0$, an interval of polar angles $\theta\in[0.1,\pi]$ and initial electron-positron momenta $|\mathbf{p}|\in[0.1,1]$~MeV.}
    \label{fig:bhabha_prop-conc-DoP_continuous}
\end{figure}

Having demonstrated a more detailed, low-level use of \texttt{QEDtool}, we now show one can use the \texttt{standard\_scattering} function. \texttt{standard\_scattering} allows for a quicker computation of all quantum-informational quantities computed by \texttt{QEDtool} for two-particle states undergoing one of the six standard QED scattering processes. However, \texttt{standard\_scattering} has the restriction of only considering scattering in the CM frame, where the incoming electron has a $3$-momentum vector parallel to the positive $z$-axis.
\begin{figure}[ht!]
    \begin{align*}
        \vcenter{\hbox{\includegraphics[width=4cm]{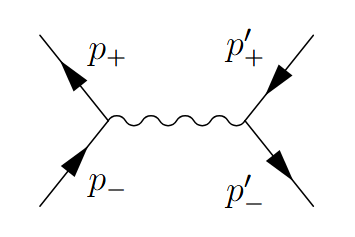}}} \quad - \quad
        \vcenter{\hbox{\includegraphics[width=3.5cm]{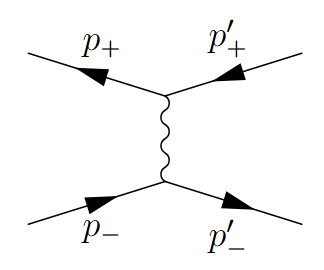}}} \\[-0.3cm]
    \end{align*}
    \caption{Leading order tree-level Feynman diagrams of $s$- and $t$-channels of Bhabha scattering. Here, $p_\pm$ and $p'_\pm$ are the initial and final positron and electron $4$-momenta, where a $-$ subscript refers to the electron and $+$ subscript refers to the positron.}
    \label{fig:Bhabha_diagrams}
\end{figure}
\begin{equation}\label{eq:bhabha_diag_in_1_2}
\begin{aligned}
    |\psi_1\rangle &= \frac{1}{\sqrt{2}}\Big( |p_-,\mathrm{L};p_+,\mathrm{L}\rangle - |p_-,\mathrm{R};p_+,\mathrm{R}\rangle \Big)\,, \\[0.2cm]
    |\psi_2\rangle &= \frac{1}{\sqrt{2}}\Big( |p_-,\mathrm{H};p_+,\mathrm{V}\rangle - |p_-,\mathrm{V};p_+,\mathrm{H}\rangle \Big)\,, \\[0.2cm] 
    |\psi_3\rangle &= |p_-,\mathrm{L};p_+,\mathrm{R}\rangle\,,
\end{aligned}
\end{equation}
[as in Eq.~(\ref{eq:convex})] equally distributed with classical weights $w_1 = w_2 = w_3 = 1/3$.
\begin{enumerate}
    \item We begin by setting the energy units to MeV and defining the electron mass ($\texttt{m}$) and charge ($\texttt{e}$) constants as shown in step 1 of Section~\ref{subsec:electron-positron}. Then, we define the arrays of initial particle $3$-momentum magnitude $\texttt{p}$ and polar scattering angle $\texttt{theta}$ values over which all quantities are to be computed. We also define an array of electron energy values:
    \begin{verbatim}
        p = np.linspace(0.1, 1, 200)
        theta = np.linspace(0.1, np.pi, 200)
        energy_e = np.sqrt(p**2 + me**2)
    \end{verbatim}
    where the lower bounds of the $|\textbf{p}|$ and $\theta$ domains are restricted in order to avoid regions of, respectively, infrared (IR) and collinear divergences. Note that azimuthal scattering angle values $\texttt{phi}$ are left undefined, because if the corresponding input of \texttt{standard\_scattering} is set to $\texttt{None}$, then all calculations are done for $\texttt{phi} = 0$.
    \item Before we implement \texttt{standard\_scattering}, we define input electron-positron momentum-helicity states, which are instances of the \texttt{QuantumState} class, as shown in step 2 of Section~\ref{subsec:electron-positron}. Here, however, we do not specify the dour-momentum label of \texttt{QuantumState}. This is due to the fact that $\texttt{standard\_scattering}$ is limited to CM scattering and does not offer the option to Lorentz-boost the scattering particles to other reference frames (unlike the rest of $\texttt{QEDtool})$.
    \begin{Verbatim}
    l = qtl.QuantumState.single(None, "L")
    r = qtl.QuantumState.single(None, "R")
    h = qtl.QuantumState.single(None, "H")
    v = qtl.QuantumState.single(None, "V")
    \end{Verbatim}
    \item The mixed input state $\rho^{\text{(in)}}$, named \texttt{in\_state}, is then defined as:
    \begin{Verbatim}
    state_1 = (l * l - r * r) / np.sqrt(2) 
    state_2 = (h * v - v * h) / np.sqrt(2)
    state_3 = l * r

    w = [1/3, 1/3, 1/3]
    states = [state_1, state_2, state_3]
        
    in_state = qtl.QuantumState.mixed(states, w)
    \end{Verbatim}
    \item \texttt{bhabha\_results} is chosen to be the name of the \texttt{pickle} file where all selected output quantities for this process are saved. We then call the \texttt{standard\_scattering} function, defining the \texttt{bhabha\_results} dictionary as
    \begin{Verbatim}
    bhabha_results = standard_scattering(in_state, "bhabha",
        p, theta, filename=None, projection=None, 
        dp=True, dcs=False, c=True, stokes=True, 
        deg_pol=True, amplitudes=False, out_state=False)
    \end{Verbatim}
    \item Calculated arrays can also be accessed directly from the list \texttt{bhabha\_results} using the respective key.
    \begin{Verbatim}
    bhabha_dp = bhabha_results["dp"]
    bhabha_c = bhabha_output["c"]
    bhabha_deg_pol = bhabha_output["deg_pol"]
    \end{Verbatim}
\end{enumerate}
Here $\texttt{bhabha\_dp}$, $\texttt{bhabha\_c}$, and $\texttt{bhabha\_deg\_pol}$ define $\partial_{\Pi}\mathcal{P}$, $C$, and $P_{(2)}$, respectively. The computed values are presented in Fig.~\ref{fig:bhabha_prop-conc-DoP_continuous}. The numerical values indicate the expected collinear and IR divergences occuring as $\theta \to 0$ and $|\mathbf{p}|\to 0$. For the in-state defined in Eq.~(\ref{eq:bhabha_diag_in_1_2}), the emitted state reaches $C \approx 1$ around $|\mathbf{p}| \approx 400$~keV for $\theta \approx \pi$. 
\begin{figure}[ht!]
    \centering
    \includegraphics{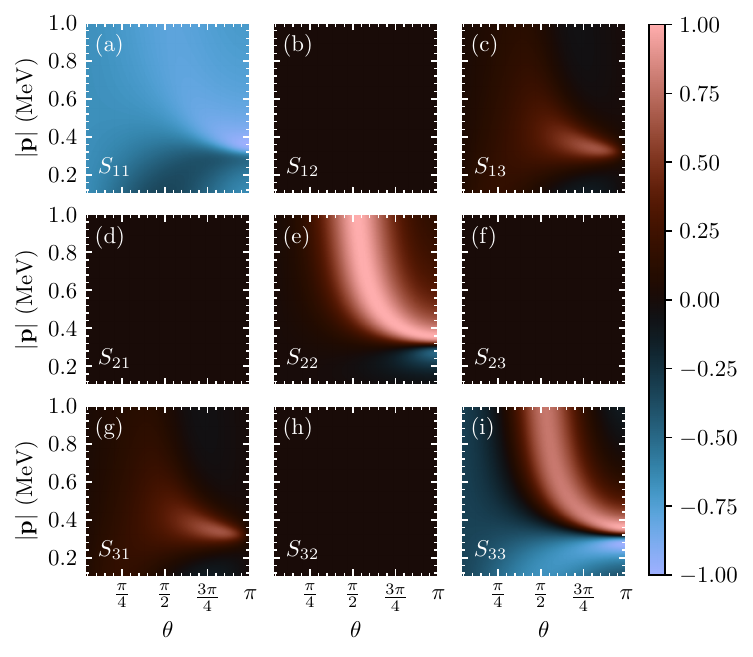}
    \caption{The Stokes parameters of the Bhabha scattered electron-positron pair. The initial electron-positron state is the classical superposition of the pure states from Eq.~(\ref{eq:bhabha_diag_in_1_2}). The results are computed in the CM frame for $\phi = 0$, an interval of polar angles $\theta\in[0.1,\pi]$ and initial electron-positron momenta $|\mathbf{p}|\in[0.1,1]$~MeV.}
    \label{fig:bhabha_Stokes}
\end{figure}

To investigate the emitted entangled state, we consider the Stokes parameters. These are obtained through
\begin{Verbatim}
    bhabha_s11 = bhabha_results["s11"]
    bhabha_s12 = bhabha_results["s12"]
    ...
\end{Verbatim}
The retrieved Stokes parameters are depicted in Fig.~\ref{fig:bhabha_Stokes}. In the regime where $C \approx 1$, we observe that $-S_{11} = S_{22} = S_{33} \approx 1$, while all other Stokes parameters are approximately zero. This corresponds to the Bell state $\smash{|\Psi^-\rangle = \big( |p'_+,\mathrm{L};p'_-,\mathrm{L}\rangle - |p'_+,\mathrm{R};p'_-,\mathrm{R}\rangle \big)/\sqrt{2}}$, as expressed in the helicity basis. This can be verified with the following lines:
\begin{Verbatim}
    >>> state = (l_e * l_p - r_e * r_p) / np.sqrt(2) 
    >>> print(qtl.stokes_parameter(state, [1, 1]),
              qtl.stokes_parameter(state, [2, 2]),
              qtl.stokes_parameter(state, [3, 3]))
    -1.0 1.0 1.0
\end{Verbatim}
Note that $|\psi_1\rangle = |\Psi^-\rangle$ [see Eq.~(\ref{eq:bhabha_diag_in_1_2})]. Thus, the $|\psi_1\rangle$ component of the in-state is likely to backscatter at $|\mathbf{p}| \approx 400$~keV, whereas $|\psi_2\rangle$ and $|\psi_3\rangle$ are not.

\section{Conclusion and outlook}\label{sec:outlook}
In this manuscript, we have introduced \texttt{QEDtool}, an open source Python package at its very first release, under the MIT license. With this version, users can perform numerical tree-level QED calculations, focus on full polarization state reconstruction, as well as entanglement and polarization correlations for both pure and mixed states. We encourage users to contribute, extend and use \texttt{QEDtool} for research and educational purposes, where applicable.

In the current release, all initial states are momentum eigenstates. In Section~\ref{subsec:S-matrix_rho}, we derived the post-scattering state for an initial state with a certain momentum wave function. A possible extension of \texttt{QEDtool} would be the implementation of this momentum wave function in the construction of the final state. Another interaction that plays a role in some technologies is the weak interaction. Currently, \texttt{QEDtool} solely encapsulates the electromagnetic interaction, and the inclusion of the weak interaction would be an enrichment to the library. Moreover, the higher the energy scale one considers, the more prominent loop diagrams and higher order corrections become. \texttt{QEDtool} currently contains no built-in functions for such renormalized corrections. Finally, as the current version contains the \texttt{FourVector} and \texttt{DiracSpinor} classes, a logical development would be to extend these to more general Lorentzian tensor and spin tensor classes.

\section*{Acknowledgments}
Jesse Smeets would like to thank P.A. Bobbert for the encouragement to pursue the development of the idea of \texttt{QEDtool}, and W.J. Holman for valuable discussions. Preslav Asenov would like to thank A. Marathe for the helpful advice regarding package licensing.

\section*{CRediT authorship contribution statement}
\textbf{Jesse Smeets:} Conceptualization, Methodology, Software, Validation, Visualization, Writing - Original Draft. \textbf{Preslav Asenov:} Software, Validation, Visualization, Writing - Review \& Editing. \textbf{Alessio Serafini:} Conceptualization, Supervision, Writing - Review \& Editing, Funding acquisition.

\section*{Funding information}
The authors acknowledge the financial support from the Leverhulme Trust Research Project Grant RPG-2024-287.

\appendix

\section{Helicity mixing coefficients}\label{ap:coeffs}

In this appendix, we will derive the helicity mixing coefficients from Section~\ref{subsec:Lorentz_transformation}. These describe general Lorentz transformations of momentum-helicity eigenstates for massive Dirac fermions. By inverting Eq.~(\ref{eq:helicity-z_relations}), we obtain
\begin{equation}\label{eq:inverted}
\begin{aligned}
    |p,\uparrow\rangle &= \exp(\i\phi_p/2)\,\Big[ \cos(\theta_p/2)\,|p,\mathrm{R}\rangle - \sin(\theta_p/2)\,|p,\mathrm{L}\rangle \Big]\,, \\[0.2cm]
    |p,\downarrow\rangle &= \exp(-\i\phi_p/2)\,\Big[ \cos(\theta_p/2)\,|p,\mathrm{L}\rangle + \sin(\theta_p/2)\,|p,\mathrm{R}\rangle \Big]\,.
\end{aligned}
\end{equation}
From Eq.~(\ref{eq:helicity-z_relations}), we find that the Lorentz transformation of a left-handed Dirac fermion, in terms of spin-$z$ eigenstates, equals
\begin{align*}
    U(\Lambda)\,|p,\mathrm{L}\rangle &= \exp(\i\phi_p/2)\cos(\theta_p/2)\,U(\Lambda)\,|p,\downarrow\rangle \\
    &\hspace{0.5cm} - \exp(-\i\phi_p/2)\sin(\theta_p/2)\,U(\Lambda)\,|p,\uparrow\rangle\,,
\end{align*}
From Eq.~(\ref{eq:spin-z_boost}), we determine $U(\Lambda)|p,M\rangle$. These transform according to
\begin{equation}\label{eq:boosted_ud_explicit}
\begin{aligned}
    &U(\Lambda)\,|p,\uparrow\rangle = \mathcal{D}_{\downarrow\uparrow}(W)\,|\Lambda p, \downarrow\rangle + \mathcal{D}_{\uparrow\uparrow}(W)\,|\Lambda p, \uparrow\rangle \,, \\[0.2cm]
    &U(\Lambda)\,|p,\downarrow\rangle = \mathcal{D}_{\downarrow\downarrow}(W)\,|\Lambda p,\downarrow\rangle + \mathcal{D}_{\uparrow\downarrow}(W)\,|\Lambda p,\uparrow\rangle \,,
\end{aligned}
\end{equation}
where the matrix elements $\mathcal{D}_{m'_s m_s}(W)$ are from Eq.~(\ref{eq:d-matrix}). We can rewrite Eq.~(\ref{eq:boosted_ud_explicit}) in terms of helicity eigenstates by using Eq.~(\ref{eq:inverted}):
\begin{equation}\label{eq:up}
\begin{aligned}
    U(\Lambda)\,|p,\uparrow\rangle &= \Big[ \mathcal{D}_{\downarrow\uparrow}(W)\exp(-\i\phi_{\Lambda p}/2) \cos(\theta_{\Lambda p}/2) \\
    &\hspace{0.8cm} - \mathcal{D}_{\uparrow\uparrow}(W)\exp(\i\phi_{\Lambda p}/2)\sin(\theta_{\Lambda p}/2) \Big]\,|\Lambda p,\mathrm{L}\rangle \\
    &\hspace{0.5cm} + \Big[ \mathcal{D}_{\uparrow\uparrow}(W)\exp(\i\phi_{\Lambda p}/2) \cos(\theta_{\Lambda p}/2) \\
    &\hspace{1.2cm}+ \mathcal{D}_{\downarrow\uparrow}(W)\exp(-\i\phi_{\Lambda p}/2)\sin(\theta_{\Lambda p}/2)\Big]\,|\Lambda p,\mathrm{R}\rangle\,,
\end{aligned}
\end{equation}
and
\begin{equation}\label{eq:down}
\begin{aligned}
    U(\Lambda)\,|p,\downarrow\rangle &= \Big[ \mathcal{D}_{\downarrow\downarrow}(W)\exp(-\i\phi_{\Lambda p}/2) \cos(\theta_{\Lambda p}/2) \\
    &\hspace{0.8cm} - \mathcal{D}_{\uparrow\downarrow}(W)\exp(\i\phi_{\Lambda p}/2)\sin(\theta_{\Lambda p}/2)\Big]\,|\Lambda p,\mathrm{L}\rangle \\
    &\hspace{0.5cm} + \Big[\mathcal{D}_{\uparrow\downarrow}(W)\exp(\i\phi_{\Lambda p}/2) \cos(\theta_{\Lambda p}/2) \\
    &\hspace{1.2cm} + \mathcal{D}_{\downarrow\downarrow}(W)\exp(-\i\phi_{\Lambda p}/2)\sin(\theta_{\Lambda p}/2)\Big]\,|\Lambda p,\mathrm{R}\rangle\,,
\end{aligned}
\end{equation}
where $\theta_{\Lambda p}$ and $\phi_{\Lambda p}$ are the spherical angles of $\Lambda p$. Consequentially, the transformation law for a left-handed electron becomes
\begin{align*}
    U(\Lambda)\,|p,\mathrm{L}\rangle &= \cos(\theta_p/2) \bigg\{ \Big[ \mathcal{D}_{\downarrow\downarrow}(W)\exp[\i(\phi_p-\phi_{\Lambda p})/2] \cos(\theta_{\Lambda p}/2) \\
    &\hspace{2.5cm} - \mathcal{D}_{\uparrow\downarrow}(W)\exp[\i(\phi_p+\phi_{\Lambda p})/2]\sin(\theta_{\Lambda p}/2)\Big]|\Lambda p,\mathrm{L}\rangle \nonumber\\[0.1cm]
    &\hspace{2.4cm} + \Big[\mathcal{D}_{\uparrow\downarrow}(W)\exp[\i(\phi_p+\phi_{\Lambda p})/2] \cos(\theta_{\Lambda p}/2) \\
    &\hspace{2.9cm}+ \mathcal{D}_{\downarrow\downarrow}(W)\exp[\i(\phi_p-\phi_{\Lambda p})/2]\sin(\theta_{\Lambda p}/2)\Big]|\Lambda p,\mathrm{R}\rangle \bigg\} \\
    &\hspace{0.5cm}- \sin(\theta_p/2) \bigg\{ \Big[ \mathcal{D}_{\downarrow\uparrow}(W)\exp[-\i(\phi_p+\phi_{\Lambda p})/2] \cos(\theta_{\Lambda p}/2) \\
    &\hspace{2.9cm}- \mathcal{D}_{\uparrow\uparrow}(W)\exp[\i(\phi_{\Lambda p}-\phi_p)/2]\sin(\theta_{\Lambda p}/2) \Big]\,|\Lambda p,\mathrm{L}\rangle \nonumber\\[0.1cm]
    &\hspace{2.8cm} + \Big[ \mathcal{D}_{\uparrow\uparrow}(W)\exp[\i(\phi_{\Lambda p}-\phi_p)/2] \cos(\theta_{\Lambda p}/2) \\
    &\hspace{3.3cm}+ \mathcal{D}_{\downarrow\uparrow}(W)\exp[-\i(\phi_p+\phi_{\Lambda p})/2]\sin(\theta_{\Lambda p}/2)\Big]|\Lambda p,\mathrm{R}\rangle \bigg\} \\[0.2cm]
    &\equiv \Upsilon_\mathrm{L}(\Lambda,p)\,|\Lambda p,\mathrm{L}\rangle + \Upsilon_\mathrm{R}(\Lambda,p)\,|\Lambda p,\mathrm{R}\rangle\,,
\end{align*}
where we defined the mixing coefficients
\begin{equation}
    \begin{aligned}
        \Upsilon_\mathrm{L}(\Lambda,p) &\equiv \cos(\theta_p/2)\Big[ \mathcal{D}_{\downarrow\downarrow}(W)\exp[\i(\phi_p-\phi_{\Lambda p})/2] \cos(\theta_{\Lambda p}/2) \\
        &\hspace{2.4cm}- \mathcal{D}_{\uparrow\downarrow}(W)\exp[\i(\phi_p+\phi_{\Lambda p})/2]\sin(\theta_{\Lambda p}/2) \Big] \\
        &\hspace{0.5cm}- \sin(\theta_p/2)\Big[ \mathcal{D}_{\downarrow\uparrow}(W)\exp[-\i(\phi_p+\phi_{\Lambda p})/2] \cos(\theta_{\Lambda p}/2) \\
        &\hspace{2.8cm}- \mathcal{D}_{\uparrow\uparrow}(W)\exp[\i(\phi_{\Lambda p}-\phi_p)/2]\sin(\theta_{\Lambda p}/2) 
        \Big]\,, \\[0.2cm]
        \Upsilon_\mathrm{R}(\Lambda,p) &\equiv \cos(\theta_p/2)\Big[\mathcal{D}_{\uparrow\downarrow}(W)\exp[\i(\phi_p+\phi_{\Lambda p})/2] \cos(\theta_{\Lambda p}/2) \\
        &\hspace{2.4cm}+ \mathcal{D}_{\downarrow\downarrow}(W)\exp[\i(\phi_p-\phi_{\Lambda p})/2]\sin(\theta_{\Lambda p}/2)\Big] \\
        &\hspace{0.5cm} - \sin(\theta_p/2)\Big[ \mathcal{D}_{\uparrow\uparrow}(W)\exp[\i(\phi_{\Lambda p}-\phi_p)/2] \cos(\theta_{\Lambda p}/2) \\
        &\hspace{2.8cm}+ \mathcal{D}_{\downarrow\uparrow}(W)\exp[-\i(\phi_p+\phi_{\Lambda p})/2]\sin(\theta_{\Lambda p}/2)\Big] \bigg\}\,.
    \end{aligned}
\end{equation}
In a similar fashion, the mixing coefficients for right-handed helicity eigenstates can be written.




\bibliographystyle{elsarticle-num}
\bibliography{bibliography}







\end{document}